\documentclass[letterpaper,10pt,openany]{article}
\usepackage[english,activeacute]{babel}
\usepackage[a4paper]{geometry}
\usepackage{graphicx}
\usepackage[textwidth=8em,textsize=small]{todonotes}
\usepackage{amsmath}
\usepackage{amssymb}
\usepackage{natbib}
\usepackage{amsfonts}
\usepackage{color}
\usepackage{graphicx}
\usepackage{flafter}
\usepackage{url}
\usepackage{enumerate}
\usepackage{multicol}
\usepackage{multirow}
\usepackage[T1]{fontenc}
\usepackage{float}
\usepackage{subfigure}
\usepackage{rotating}
\usepackage{comment}
\usepackage{cancel}
\usepackage{booktabs}
\restylefloat{table}

\def\exp#1{{\rm exp}{#1}}
\def\frac#1#2{{{#1}\over{#2}}}

\DeclareMathOperator*{\tr}{tr}

\newcommand\simiid{\mathrel{\overset{\makebox[0pt]{\mbox{\normalfont\tiny\sffamily iid}}}{\sim}}}
\newcommand\simind{\mathrel{\overset{\makebox[0pt]{\mbox{\normalfont\tiny\sffamily ind}}}{\sim}}}

\newcommand{\expec}[1]{\textsf{E}\left[#1\right]}

\newcommand{\cov}[1]{\textsf{C}\textsf{ov}\left[#1\right]}

\newcommand{\diag}[1]{\text{diag}\left[#1\right]}

\newcommand{\ex}[1]{\exp{ \left\{ #1 \right\}}}

\def\A{\mathbf{A}}

\def\I{\mathbf{I}}

\def\mv{\boldsymbol{m}}

\def\V{\mathbf{V}}
\def\W{\mathbf{W}}
\def\X{\mathbf{X}}\def\xv{\boldsymbol{x}}
\def\yv{\boldsymbol{y}}
\def\Z{\mathbf{Z}}\def\zv{\boldsymbol{z}}
\def\al{\alpha}\def\alv{\boldsymbol{\alpha}}
\def\be{\beta}\def\bev{\boldsymbol{\beta}}

\def\eps{\epsilon}\def\epsv{\boldsymbol{\epsilon}}

\def\te{\theta}\def\tev{\boldsymbol{\theta}}

\def\sig{\sigma}

\def\epsv{\boldsymbol{\eps}}
\def\tev{\boldsymbol{\theta}}

\def\UPS{\mathbf{\Upsilon}}

\def\Xiv{\boldsymbol{\Xi}}
\def\Phiv{\boldsymbol{\Phi}}
\def\Upsv{\boldsymbol{\Upsilon}}
\def\alphav{\boldsymbol{\alpha}}
\def\betav{\boldsymbol{\beta}}

\def\Ber{\small{\mathsf{Ber}}}

\def\Nor{\small{\mathsf{N}}}

\def\IGamd{\small{\mathsf{IGam}}}

\def\trans{\text{\textsf{T}}}
\def\zerov{\boldsymbol{0}}
\def\onev{\boldsymbol{1}}

\def\reals{\mathbb{R}}
\allowdisplaybreaks

\begin{document}

\title{Time-Varying Coefficient Model Estimation Through Radial Basis Functions}

\author{
	Juan Sosa \\ 
	Universidad Nacional de Colombia, Colombia \footnote{jcsosam@unal.edu.co} \\ 
	Lina Buitrago \\
	Universidad Nacional de Colombia, Colombia \footnote{labuitragor@unal.edu.co}}

\date{}

\maketitle


\begin{center}
	\textbf{Abstract}
\end{center}

In this paper we estimate the dynamic parameters of a time-varying coefficient model through radial kernel functions in the context of a longitudinal study. Our proposal is based on a linear combination of weighted kernel functions involving a bandwidth, centered around a given set of time points. In addition, we study different alternatives of estimation and inference including a Frequentist approach using weighted least squares along with bootstrap methods, and a Bayesian approach through both Markov chain Monte Carlo and variational methods. We compare the estimation strategies mention above with each other, and our radial kernel functions proposal with an expansion based on regression spline, by means of an extensive simulation study considering multiples scenarios in terms of sample size, number of repeated measurements, and subject-specific correlation. Our experiments show that the capabilities of our proposal based on radial kernel functions are indeed comparable with or even better than those obtained from regression splines. We illustrate our methodology by analyzing data from two AIDS clinical studies.


\section{Introduction}\label{sec_introduction}

Statistical models for longitudinal data are powerful instruments of analysis when experimental units (subjects) are measured repeatedly over time in relation to a response variable along with static or time-dependent covariates. A very important feature of this kind of data that must be taken into account when fitting a statistical model, is the likely presence of serial correlation within repeated measurements on a given subject (observations between subjects are assumed to be independent). Typically, the main purpose of the analysis is to identify and characterize the evolution (mean tendency) of the response variable over time and quantify its association with the covariates. Parametric techniques for longitudinal data analysis have been exhaustively studied in the literature \citep[see for example][and references within] {molenberghs-2014, liu-2015, little-2015}. Though useful in many cases, questions about the adequacy of the assumptions of parametric models and the potential impact of model misspecification on the analysis often arise. For instance, one of the basic assumptions associated with parametric techniques, yet not always satisfied, establishes that the mean response must be a known function of both fixed and random effects, indexed by a set of unknown parameters. Thus, for many practical situations, parametric models may be too restrictive or even unavailable. 


In order to overcome such difficulties, building on the contributions of \cite{cleveland-1991}, \cite{hastie-tibshirani-1993}, and \cite{zeger-diggle-94}, the work of \cite{hoover-rice-wu-yang-98} considered a nonparametric model that lets the parameters vary over time. Nonparametric models of this nature allow more flexible functional dependence between the response variable and the covariates, since they are based on time-dependent coefficients (smooth functions of time) instead of fixed unknown parameters. Due to their interpretability and flexibility, these models have been the subject of active research over the last twenty years. 
{\color{black} Early popular developments are given in \cite{wu-1998-asymptotic}, \cite{zhang-1998-semiparametric}, \cite{fan-zhang-1999-statistical}, \cite{wu-2000-kernel}, \cite{cai-2000-efficient}, \cite{fan-2000-two}, \cite{lin-2000-nonparametric}, \cite{lin-2001-semiparametric}, \cite{chiang-2001-smoothing}, \cite{rice-2001-nonparametric}, \cite{wu-2002-local}, and \cite{huang-2002-varying}. For applications and surveys, see \cite{fan-2008-statistical}, \cite{tan2012time}, \cite{zhang-2013}, and \cite{wu-tian-2018}.}

Specifically, consider the longitudinal dataset 
$$
{\color{black} \mathcal{D} =} \left\{(y_{i,j}, \xv_{i,j}, t_{i,j}, ) : j=1,\ldots,n_i, i=1,\ldots,n\right\}\,,
$$ 
where $y_{i,j} \equiv y_i(t_{i,j})$ and $\xv_{i,j} \equiv \xv_i(t_{i,j})$ with $\xv_i(t) = \left(x_{0,i}(t),x_{1,i}(t),\ldots,x_{d,i}(t)\right)$ are the real-valued response variable and the $(d+1)$ column covariate vector,  corresponding to measurement $j$ of subject $i$, observed at time $t_{i,j}$, $n$ is the number of subjects, $n_i$ is the number of observations associated with subject $i$, and the total number of observations in the sample is $N = \sum_{i=1}^n n_i$. Measurement times are often distinct and irregularly spaced in a fixed interval of finite length. In order to evaluate the mean joint effect of time $t$ and the covariates $\xv(t)$ on the outcome $y(t)$, we use the structured nonparametric model
\begin{equation}\label{eq_tvcm}
y(t) = \xv(t)^\trans\bev(t) + \epsilon(t) \\
\end{equation}
where $\bev(t) = (\beta_{0}(t),\beta_{1}(t),\ldots,\beta_{d}(t))$ is a $(d+1)$ column vector of real-valued nonparametric functions of time $t$, called dynamic coefficients or dynamic parameters, and $\epsilon(t)$ is a zero-mean stochastic process, independent of $\xv(t)$, with covariance function $\gamma(s,t) = \cov{\eps(s), \eps(t)}$. The model in equation \eqref{eq_tvcm} is referred to as a {\color{black} (fixed-effects)} time-varying coefficient model (\textsf{TVCM}). This model provides a parsimonious approach for characterizing time varying association patterns of a set of dynamic predictors on the expected value of a functional response. Notice that this model has a natural interpretation since for a fixed time point $t$, the \textsf{TVCM} \eqref{eq_tvcm} reduces to a multiple linear model with response variable $y(t)$ and covariate vector $\xv(t)$, for which a standard interpretation of the time-varying coefficients $\be_r(t)$, $r = 0,1,\ldots,d$, holds. In our experiments, we take $x_0(t) \equiv 1$ for all $t$, which means that $\beta_0(t)$ is the intercept coefficient describing the baseline time-trend. {\color{black} Finally, note that a discretized version of the \textsf{TVCM} can be obtain by simply substituting $t$ by $t_{i,j}$ in equation \eqref{eq_tvcm}, considering all data points given in $\mathcal{D}$, in order to highlight the dependence on data when necessary.}

{\color{black} In order to illustrate the full potential that a model such as a \textsf{TVCM} has, we provide below a fair revision on extensions of the model that have taken place over recent years.} Motivated by applications rather than simply a desire to modify a statistical model, several extensions of the \textsf{TVCM} \eqref{eq_tvcm} have been proposed over the years in all sorts of directions, along with more complex  structures and weaker assumptions. Some of these extensions typically share characteristics with each other. Here, we list some relevant instances in no particular order. A popular extension emerged naturally in order to efficiently capture both population and individual relationships. In that way, mixed-effects time-varying coefficient models extended \textsf{TVCM}s by dividing the error term into two parts, one part representing the subject-specific deviation from the population mean function, and the other representing the measurement error \citep{liang-2003-relationship, wu-2004-backfitting, lu-2009-smoothing, sosa-2012-random, chiou-2012-functional, jeong-2016-bayesian}; {\color{black} we consider a model of this sort in our first simulation study (see Section \ref{sec_simulation_1} for details).} Another widespread extension took place when non-Gaussian responses where modeled directly, ranging from dichotomous and categorical outcomes to variables with skewed distributions, which provided a unified framework to do so. Thus, generalized time-varying coefficient models extended \textsf{TVCM}s by introducing a known link function to relate the dynamic linear predictor and the response process to each other \citep{biller-2001-bayesian, cai-2000-efficient, csenturk-2008-generalized, lu-2009-smoothing,  csenturk-2013-modeling, lu-2017-bayesian, jeong-2017-analysis}. Also, more complex types of dynamic functional dependencies have been developed, such as the relationships provided in time-varying additive and nonlinear models \citep{fan-2003-adaptive, qu-2006-quadratic, wang-2007-varying, csenturk-2010-functional, wu-2013-nonparametric}. In addition, more adaptations were developed to deal with the same issues that non-varying models deal with. For instance, quantile regression \citep{wang-2009-quantile,andriyana-2018-quantile}, variable selection and shrinkage estimation \citep{fan-2005-profile,li-2008-variable,wang-2008-variable,wang-2009-shrinkage,wei-2011-variable}, and even spatial modeling \citep{assunccao-2003-space,gelfand-2003-spatial,waller-2007-quantifying,wu-2010-nonparametric,serban-2011-space,nobles-2014-spatial,jeong-2016-bayesian}.

On the other hand, several smoothers can be used to estimate the dynamic coefficients of \textsf{TVCM} \eqref{eq_tvcm}. The key idea behind the estimation process relies on rewriting each $\be_r(t)$ through a linear expansion of parametric functions in order to make possible parametric-like inference as in standard models. In general, each smoother is indexed by a smoothing parameter vector that controls the trade-off between goodness-of-fit and model complexity. Thus, smoothing parameter selection criteria are in order. Some of the most popular smoothers include local polynomial smoothers, regression spline, smoothing spline, and P-splines \citep{wu-zhang-06,wu-tian-2018}. Different smoothers have different strengths in one aspect or another. For example, smoothing spline may be good for handling sparse data, while local polynomial smoothers may be computationally advantageous for handling dense designs.

The purpose of this paper is twofold. First, in order to estimate the time-varying coefficients $\beta_r(t)$, we propose a linear smoother based on kernel functions, by treating them as if they were radial basis functions \citep[see][for a complete catacterization of radial basis funcions]{buhmann-04}. This approach has been used in both the semiparametric regression \citep{ruppert-2003-semiparametric} and statistical learning literature \citep{hastie-2009-elements,harezlak-2018-semiparametric} but, to the best of our knowledge, it has not been fully exploited yet in the context of longitudinal data analysis, apart from the work of \cite{serban-2011-space} on space-time varying coefficients models. Our proposal applies to both time-invariant and time-dependent covariates as well as regular and irregular placed design times, and also allows for different amounts of smoothing for different coefficients. Second, since a radial kernel expansion resembles very closely an approximation using spline basis functions, we compare these smoothing alternatives with each other in terms of goodness-of-fit and prediction using both Frequentist and Bayesian inference frameworks. To that end, from a Frequentist perspective, we consider weighted least squares and bootstrap methods. From a Bayesian perspective, we consider Markov chain Monte Carlo (MCMC) along with variational methods. The Bayesian approach has become more popular in recent years  \citep{biller-2001-bayesian, waller-2007-quantifying, hua-2011-bayesian, memmedli-2012-application, jeong-2016-bayesian, lu-2017-bayesian, franco-2019-unified}, but variational algorithms have not been explored to ease the computational burden under this framework.

The remainder of the paper is structured as follows:
Section \ref{sec_estimation} introduces the estimation of time-varying coefficients through radial kernel functions and regression spline. Section \ref{sec_computation} discusses different approaches to statistical inference. Section \ref{sec_selection_number_location_knots} discusses the choice of knots and smoothing parameter selection. Section \ref{sec_simulation} compares the estimation alternatives through an extensive simulation study. Section \ref{sec_illustrations} illustrates our proposal by analyzing AIDS data coming from two clinical studies. Finally, Section \ref{sec_discussion} presents some concluding remarks and directions for future work.

\section{Estimation using radial kernel functions}\label{sec_estimation}

The main idea behind estimation through radial kernel functions consists in expressing each dynamic coefficient in model \eqref{eq_tvcm} as a linear combination of kernel functions by treating them as radial basis functions. A radial smoother can be constructed using the following set of radial basis:
\begin{equation}\label{eq_radial_basis}
1, t, \ldots, t^g, \xi(|t-\kappa_1|),\ldots,\xi(|t-\kappa_k|)\,,
\end{equation}
where $\xi(\cdot)$ is a kernel function, $|\cdot|$ is the Euclidean norm in $\reals$, and $\kappa_1<\ldots<\kappa_k$ are $k$ knots covering the time domain. The smoother performance strongly depends on the proper selection of both location and number of knots (see Section \ref{sec_selection_number_location_knots} for details). The basis degree $g$ is usually less crucial and it is typically taken as 1, 2, or 3, for computational convenience. On the other hand, note that the first $g + 1$ basis functions of \eqref{eq_radial_basis} are polynomials of degree up to $g$, and the others are all kernel functions, which satisfy the property $\xi(t) = \xi(|t|)$. Such functions are known as radial functions \citep{buhmann-04}. Different kinds of kernel functions are commonly used in practice, such as Gaussian or Epanechnikov kernels, among many others \citep[see][for a review]{wasserman-06}. This choice is less significant in terms of smoothing.

Using the radial basis \eqref{eq_radial_basis}, we can express each time-varying coefficient $\be_r(t)$, $r=0,1,\ldots,d$, as
\begin{equation}\label{eq_radial_basis_expansion}
\beta_r(t)=\sum_{\ell=0}^{g}\alpha_{r,\ell}\,t^\ell + \sum_{\ell=1}^{k_r} \alpha_{r,g+\ell}\,\xi(|t-\kappa_\ell|) = \Xiv_r(t)^{\trans}\alv_r\,,
\end{equation}
where 
$\Xiv_r(t) = \left(1, {\color{black} t,} \ldots, t^g, \xi(|t-\kappa_1|),\ldots,\xi(|t-\kappa_{k_r}|)\right)$
and 
$\alv_r = (\al_{r,0},\ldots,\al_{r,k_r+g+1})$ 
are $p_r\times 1$ column vectors with $p_r=k_r+g+1$, composed of basis functions evaluated at time $t$ and unknown parameters, respectively. Such a representation is able to accommodate a variety of shapes and smoothness for the dynamic coefficients without overfitting the data, since separate number of knots are allowed. This means that, for a fixed degree basis $g$, the number of knots $k_0,\ldots,k_d$ play the role of smoothing parameters.

In this way, the dynamic vector $\bev(t)$ in model \eqref{eq_tvcm} becomes
\begin{equation}\label{eq_radial_basis_expansion_2}
\bev(t)=\Xiv(t)^\trans \alv\,,
\end{equation}
where $\alv=(\alv_0^\trans,\ldots,\alv_d^\trans)$ and $\Xiv(t)=\text{diag}[\Xiv_{0}(t),\Xiv_{1}(t),\ldots,\Xiv_{d}(t)]$. Note that $\alv$ is a column vector of size $p\times 1$, $p=\sum_{r=0}^d p_r$, whereas $\Xiv(t)$ is a rectangular matrix of size $p\times (d+1)$. Now, substituting $\bev(t)$ in model \eqref{eq_tvcm} for its equivalent expression given in \eqref{eq_radial_basis_expansion_2}, it follows that the \textsf{TVCM} \eqref{eq_tvcm} can be approximately written as
\begin{equation}\label{eq_TVCM_component_form}
\begin{gathered}
y_{i,j} = \zv_{i,j}^\trans\alv + \eps_{i,j}\,,\qquad
j=1,\ldots,n_i, \,\qquad i=1,\ldots,n,
\end{gathered}
\end{equation}
where $y_{i,j} \equiv y_i(t_{i,j})$, $\eps_{i,j} \equiv \eps_i(t_{i,j})$, and $\zv_{i,j} = (\zv_{0,i,j}^\trans,\ldots,\zv_{d,i,j}^\trans)$, with 
\begin{equation}\label{eq_z_rij}
\zv_{r,i,j}=x_{r,i}(t_{i,j})\,\Xiv_{r}(t_{i,j})\,\qquad r = 0,1,\ldots,d\,.
\end{equation} 
Similar to $\alv$, each $\zv_{i,j}$ is a $p\times 1$ column vector of covariate values times radial basis functions. For the $i$-th subject, $i=1,\ldots,n$, we denote the response vector, the random error vector, and the design matrix as
$$
\yv_i   = (y_{i,1},   \ldots,y_{i,n_i})   \,, \qquad 
\epsv_i = (\eps_{i,1},\ldots,\eps_{i,n_i})\,, \qquad  
\Z_i    = [\zv_{i,1}, \ldots,\zv_{i,n_i}]^\trans\,.
$$
Consistently, we denote the response vector, the random error vector, and the design matrix  for the whole dataset as 
$$
\yv   = (\yv_1^\trans,\ldots,\yv_n^\trans),\,\, \qquad
\epsv = (\epsv_1^\trans,\ldots,\epsv_n^\trans),\,\, \qquad 
\Z    = [\Z_1^\trans, \ldots,\Z_n^\trans]^\trans,\,\, 
$$
which allow us to express model \eqref{eq_TVCM_component_form} in matrix form as a standard linear model:
\begin{equation}\label{eq_TVCM_matrix_form}
\yv=\Z\alv + \epsv\,.
\end{equation}
Once an estimate for $\alphav$, $\widehat{\alphav}=(\widehat{\alphav}_0^\trans,\ldots,\widehat{\alphav}_d^\trans)$, is obtained; it is straightforward to get an estimate for $\betav(t)$, $\widehat{\betav}(t)=(\widehat{\beta}_0(t),\ldots,\widehat{\beta}_d(t))$, by simply letting $\widehat{\betav}(t)= \Xiv(t)^\trans \widehat{\alphav}$. Therefore, our task reduces to estimate $\alv$ in model \eqref{eq_TVCM_matrix_form} using an appropriate method.

Similarly, the key concept working with spline functions is to represent each dynamic coefficient through a regression spline basis, such as the truncated power basis, B-spline basis, or wavelet basis, among others \citep[see][for a review]{ramsay-2009-functional}. {\color{black} The B-spline basis} is a powerful choice due to its simplicity and capability to capture local features of dynamic relationships. In this way, emulating the methodology described above, in order to estimate the time-varying coefficient $\beta_r(t)$, we consider the following expansion based on truncated power functions:
\begin{equation}\label{eq_truncated_power_basis_expansion}
\beta_r(t)=\sum_{\ell=0}^{g}\alpha_{r,\ell}\,t^\ell + \sum_{\ell=1}^{k_r} \alpha_{r,g+\ell}\, (t-\kappa_{\ell})^g_{+}\, = \Phiv_r(t)^{\trans}\alv_r\,,
\end{equation}
where $x_{+}^g$ denotes the $g$-th power of the positive part of $x$, $x_{+}=\max(0,x)$, and $\kappa_1,\ldots,\kappa_{k_{r}}$ are $k_r$ knots (in an increasing order) scattered in the range of interest. As before, note that $\Phiv_r(t) = \left(1, {\color{black} t,} \ldots, t^g, (t-\kappa_1)^g_{+},\ldots,(t-\kappa_{k_r})^g_{+}\right)$ and $\alv_r = (\al_{r,0},\ldots,\al_{r,k_r+g+1})$ are $p_r\times 1$ column vectors, $p_r=k_r+g+1$, composed of basis functions evaluated at time $t$ and unknown parameters, respectively. Then, it is simple to obtain an estimate for $\bev(t)$ as $\widehat{\betav}(t)= \Phiv(t)^\trans \widehat{\alphav}$, where $\Phiv(t)=\text{diag}[\Phiv_{0}(t),\Phiv_{1}(t),\ldots,\Phiv_{d}(t)]$ and $\widehat\alv$ is an estimate of $\alv$ in model \eqref{eq_TVCM_matrix_form}, whose design matrix $\Z$ is constructed from \eqref{eq_z_rij} using $\Phiv_r(t_{i,j})$ instead of $\Xiv_r(t_{i,j})$.

{\color{black}
Finally, the reader should note that our proposal given in equation \eqref{eq_radial_basis_expansion} is a direct reformulation of the expansion based on truncated power functions (which have been extensively investigated in the literature as in \citealp{wu-zhang-06}, for example; and therefore, we use them as a baseline), obtained by using other kind of basis functions.  We argue that this is a sensible thing to do since radial functions (kernels in particular) have very desirable properties (e.g. the semiparametric regression expansion considered here based on radial functions are kernel machines within the reproducing kernel Hilbert space framework; see \citealp{harezlak-2018-semiparametric}, for example) to representing (smoothing) all sorts of functional behaviors. That is why we believe that our approach constitutes a reasonable choice to represent dynamic coefficients in any \textsf{TVCM}.  As a final comment, we note that, within a given inference paradigm, the ``computational complexity'' of either expansion is equivalent, because each basis is composed of $1+g+K$ real-valued functions.
}

\section{Inference methods}\label{sec_computation}

According to the previous section, \textsf{TVCM} \eqref{eq_tvcm} is locally equivalent to standard linear model \eqref{eq_TVCM_matrix_form} in which it is required to estimate the parameter vector $\alv$.  {\color{black} In what follows, we consider both Frequentist and Bayesian approaches to carry out statistical inference on $\alv$, and as a consequence, on $\betav(t)$. First, we motivate a very popular sampling distribution as well as widely known classical bootstrap methods for performing statistical inference. Then, we consider information external to the dataset by means of a conjugate prior distribution, along with our proposal for quantifying uncertainty based on simulation and variational methods. We discuss in detail implications, challenges, and algorithms for each protocol, but focus our attention on our estimation approach embedded in the Bayesian paradigm.}

\newpage

\subsection{Frequentist inference}

{\color{black}
From a likelihood point of view, in its simplest form, we can consider the sampling distribution
\begin{equation}\label{eq_y_sampling_distribution}
\yv  \mid \Z,\W,\alv,\sig^2 \sim \Nor(\Z\alv,\sig^2\W^{-1})\,,
\end{equation}
where $\W=\diag{\W_1,\ldots,\W_n}$ and $\W_i = w_i\I_{n_i}$ is the weight matrix for the $i$-th experimental unit, $i = 1,\ldots,n$, which is equivalent to assuming $\epsv\mid\W,\sig^2\sim\Nor(\zerov,\sig^2\W^{-1})$ in model \eqref{eq_TVCM_matrix_form}, in a way that $\eps(t) \sim \text{\textsf{GP}}(\mu, \gamma)$ is a Gaussian process with $\mu(t)=0$ and $\gamma(s,t) = \sigma^2\, 1_{\{ s = t\}}$ in model \eqref{eq_tvcm}.  The weights $w_1,\ldots,w_n$ are known positive constants such that $\sum_{i=1}^n n_i w_i = 1$, which quantify the relative importance of experimental units. In our experiments, we follow \cite{wu-tian-2018} and consider the ``subject uniform weight'', $w_i =1/(nn_i)$, where each subject is inversely weighted by its number of repeated measurements $n_i$, so that the subjects with fewer repeated measurements receive more weight than the subjects with more repeated measurements.  The above independence assumption is convenient mathematically and works well when longitudinal data tend to be sparse.  However, in our experience, and also, considering empirical evidence from both \cite{wu-zhang-06} and \cite{wu-tian-2018}, such an assumption can be robust to some deviations from data sparsity. Therefore, the sampling distribution \eqref{eq_y_sampling_distribution} is an appealing choice in practice.

Under this setting, the resulting maximum likelihood estimator of $\alv$ is given by
\begin{equation}\label{eq_WLS_alpha_estimator}
\widehat\alv = (\Z^\trans\W\,\Z)^{-1}\Z^\trans\W\yv\,,
\end{equation}
which is equivalent to the estimator obtained as a result of minimizing the weighted least squares (\textsf{WLS}) criterion
\begin{equation}\label{objective_function_WLS}
\textsf{WLS}(\alphav) = \sum_{i=1}^{n}\sum_{j=1}^{n_i}w_i (y_{i,j} - \zv_{i,j}^\trans\,\alphav)^2\,.
\end{equation}
Note that according to the Gauss-Markov theorem, the estimator provided in \eqref{eq_WLS_alpha_estimator} is the best linear unbiased estimator (BLUE) of $\alv$. Furthermore, it can be shown that an unbiased estimator of $\sig^2$ is
\begin{equation}\label{eq_sig2_estimator}
\widehat\sig^2 = \frac{1}{N-p}\, (\yv - \Z\widehat\alv)^\trans\W\, (\yv - \Z\widehat\alv)\,,
\end{equation}
where $N = \sum_{i=1}^n n_i$ is the total number of observations, $p=\sum_{r=0}^d p_r$ is the expansion dimension, and $\widehat\alv$ is the estimator of $\alv$ given in \eqref{eq_WLS_alpha_estimator}.
}

Under the Frequentist paradigm, confidence intervals can be computed based on either asymptotic distributions or bootstrap methods \citep{efron-2016-computer}. However, given the compound structure of longitudinal data, inferences that are based on asymptotic distributions are typically difficult to justify in practice, since they heavily rely on assumptions that are difficult to meet. Thus, we consider bootstrap methods, which can always be implemented based on the available data regardless of sample sizes and sampling distributions. See Appendix \ref{app_bootstrap} for a detailed description of the bootstrap algorithm. The main advantage of using a bootstrap procedure is that it does not rely on asymptotic distributions and can be used to construct confidence intervals. For instance, at a given time $t$, the $100(1-\alpha)\%$ percentile-based confidence interval for $\be_r(t)$, $r=0,1,\ldots,d$, is given by 
\begin{equation}\label{eq_IC_betas}
\left( \be_{r,\alpha/2}(t), \be_{r,1-\alpha/2}(t) \right)\,,
\end{equation}
where $\be_{r,\alpha/2}(t)$ and $\be_{r,1-\alpha/2}(t)$ are the $\al/2$ and $1-\alpha/2$ quantiles of the bootstrap samples $\be_r(t)^{(1)},\ldots,\be_r(t)^{(B)}$, which are computed based on $\alv^{(1)},\ldots,\alv^{(B)}$ and a given set of basis functions.
Other types of confidence intervals are available \citep[e.g., normal-based confidence intervals; see][for a review]{efron-2016-computer}. {\color{black} We highlight that the confidence intervals given above correspond to pointwise confidence sets that only work for $\be_r(t)$ at a given time $t$. In most practical situations, such pointwise inferences are sufficient. However, in some studies, we might require a confidence band that simultaneously includes the true dynamic coefficient $\be_r(t)$ for a range (typically large) of time values. In such situations, we need to construct a simultaneous confidence band for $\be_r(t)$ for $t$ within a given time interval. We refer the reader to \cite{wu-tian-2018} for details about this matter. 
	
Even though theoretical properties of bootstrap procedures in this setting have not been systematically investigated, we are quite confident about the coverage rates in this case, given previous simulations studies about this matter as in \cite{hoover-rice-wu-yang-98} and \cite{wu-2000-kernel} \cite[see also][and refereces therein for a comprehensive review and also more empirical evidence in this regard]{wu-tian-2018}. 
}

\subsection{Bayesian inference}\label{sec_bayesian_inference}

Under a Bayesian framework, in order to obtain an estimate for $\alv$ under the sampling distribution \eqref{eq_y_sampling_distribution}, it suffices to consider the standard normal regression model  
\begin{equation}\label{eq_standard_linear_model}
\yv\mid\Z,\alv,\sig^2\sim\Nor(\Z\alv,\sig^2\I_N)\,,
\end{equation}
since it can be easily obtained from \eqref{eq_y_sampling_distribution} by means of a linear transformation on $\yv$ based on a Choleski factorization of $\W^{-1}$ \citep[see][for details]{faraway-2014-linear}. In what follows, we consider the sampling distribution \eqref{eq_standard_linear_model}, having in mind that a preprocessing step is required before fitting the model. In order to complete the model we choose the so-called independent Zellner's $\textsf{g}$-prior as a simple semiconjugate prior distribution on $\alv$ and $\sigma^2$ to be used when there is little prior information available. Under this invariant $g$-prior, we let
$$
\alv \mid \sig^2 \sim \Nor(\zerov,N\sig^2\I_N)
\qquad\text{and}\qquad
\sig^2 \sim \IGamd(a_\sig,b_\sig)\,,
$$
where $a_\sig$ and $b_\sig$ are known hyperparameters.

Regarding the hyperparameter elicitation, we need a prior distribution to be as minimally informative as possible in the absence of real external information. We recommend setting $\textsf{g} = N$, $a_\sig = 2$, and $b_\sig = \widehat\sigma^2$ as in \eqref{eq_sig2_estimator}. This choice of $\textsf{g}$ makes $\tfrac{\textsf{g}}{\textsf{g}+1}$ very close to 1, and therefore, we are practically centering $\alv$ around $\widehat\alv$ a priori. Similarly, the prior distribution of $\sigma^2$ is also weakly centered around $\widehat\sig^2$, since $a_\sig=2$ implies an infinite variance on $\sig^2$ a priori.  Such a distribution cannot be strictly considered as a real prior distribution, as it requires knowledge of $\yv$ to be constructed. However, it only uses a small amount of the information in $\yv$, and can be loosely thought of as the prior distribution of a researcher with unbiased but weak prior information \citep[see][for a discussion]{hoff-2009-first}. Refer to Appendix \ref{app_mcmc} for details about the Gibbs sampler.

Even though the MCMC algorithm is straightforward in this case, inference may become impractical as the number of experimental units and the number of covariates grow. For this reason, we also implement a variational Bayes alternative that can potentially alleviate the computational burden in big data scenarios. See Appendix \ref{app_variational} for details regarding the variational algorithm.

\section{Location and number of knots}\label{sec_selection_number_location_knots}

{\color{black} The smoothers'} quality strongly depends on both knot locations and the number of knots. The degree of the expansion $g$ is usually less crucial and it is often taken as 1, 2, or 3. In terms of knot location, we distinguish two widely-used alternatives. The first method locates equally spaced points in the range of interest, independently of the design time points. It is usually employed when the design time points are uniformly scattered in the range of interest. The second method locates equally spaced quantiles of the design time points as knots. It locates more knots where more design time points are scattered. These methods are equivalent when the design time points are uniformly scattered.  {\color{black} However, in our experience, the equally spaced method to locate knots is very convenient due to both its simplicity and proclivity to work well even in all sort of situations.}

Another essential feature that we need to handle in practice is how to choose the smoothing parameter vector $\boldsymbol{p} = (p_0,\ldots,p_d)$. A popular method to do so is the so called leave-one-point-out cross-validation \cite[\textsf{PCV},][]{eubank-et-al-04}. This approach aims to select a good smoothing parameter vector via trading-off the goodness-of-fit and the model complexity. The idea behind this criteria consists in choosing the smoothing parameter vector $\boldsymbol{p}$ that minimizes the expression
\begin{equation}\label{eq_PCV_original_formula}
\textsf{PCV} = \sum_{i=1}^n \sum_{j=1}^{n_i} w_i \left[ y_{i,j} - \xv_i(t_{i,j})^\trans\widehat{\betav}^{(-i,j)}(t_{i,j}) \right]^2\,,
\end{equation}
where $\widehat{\betav}^{(-i,j)}(t_{i,j})$ is an estimate of $\betav(t_{i,j})$ using the entire dataset except the $j$-th measurement of the $i$-th experimental unit. It can be shown that expression \eqref{eq_PCV_original_formula} is equivalent to
\begin{equation}\label{PCV_smooth_matrix}
\textsf{PCV} = \frac{(\yv-\A\yv)^\trans\W(\yv-\A\yv)}{\left(1-\text{tr}(\A)/N\right)^2}\,,
\end{equation}
where $\text{tr}(\A)$ is the trace of the smoothing matrix $\A$, which is a square matrix such that $\widehat\yv = \A\yv$. 
Even though the \textsf{PCV} criteria does not account for the within-subject correlation effectively, it is a suitable method since the computational performance is substantially better than that provided by other alternatives \citep[e.g., leave-one-subject-out cross-validation; see][for details]{wu-tian-2018}. As discussed in Section \ref{sec_discussion}, other alternatives relying on model-based knot introduction or deletion are available.

\section{Simulation Study} \label{sec_simulation}

In this section, we present two benchmark simulation scenarios to evaluate the performance of our proposed methodology and compare the different inferential methods. The first simulation scenario is inspired on an experiment originally proposed by \cite{wu-2004-backfitting} and \cite{wu-zhang-06}. The second experiment follows very closely a simulation study performed by \cite{wu-2000-kernel}, \cite{huang-2002-varying}, and \cite{wu-tian-2018}.

\subsection{Simulation scenario 1} \label{sec_simulation_1}

{\color{black} In order to test our methodology with challenging real-life like datasets, and also, evaluating the robustness of the model to typical deviations from the true data generating process, we consider in this experiment a mixed-effects time-varying coefficient model with no covariate information. Such a model takes a (fixed-effects) \textsf{TVCM} with $d=0$ and $x_0(t)\equiv 1$ for all $t$, and decomposes the error term into two random parts: the first one, which is subject-specific, describes the characteristics of each individual that deviate from the mean population behavior; whereas the second one, which handles directly pure random error, encompasses all those factors out of reach by the modeler (such as error measurement). Thus,} we generate synthetic datasets as follows:
\begin{equation}\label{eq_tvcm_sim1}
y_i(t) = \beta_0(t) + \upsilon_{i}(t) + \epsilon_i(t)\,,\qquad i = 1,\ldots,n,
\end{equation}
where $\beta_0(t)$ is a known time-varying coefficient, $\upsilon_{i}(t) =a_{i,0} + a_{i,1}\cos(2 \pi t) + a_{i,2}\sin(2 \pi t)$ is a subject-specific random effect, with $(a_{i,0},a_{i,1},a_{i,2})\simiid \Nor\left(\zerov,\text{diag}[\sigma_0^2,\sigma_1^2,\sigma_2^2]\right)$, and $\epsilon_i(t)$ is a zero-mean Gaussian process such that $\epsilon_i(t)\simind\Nor(0,\sigma_\epsilon^2 [1-\text{exp}(-0.5t - i/n)]^2)$. We assume that $\sigma_1^2=\sigma_2^2=\sigma_\epsilon^2=\sigma^2$, and therefore, the correlation between repeated measurements $\rho$ within each experimental unit is bounded by $(\sigma_0^2+2\sigma^2)^{-1}(\sigma_0^2-\sigma^2)$ and $(\sigma_0^2+2\sigma^2)^{-1}(\sigma_0^2+\sigma^2)$. We consider three cases in order to simulate different correlation levels, namely, weak within-subject correlation, $\sigma^2=0.01$ and $\sigma_0^2=0.01$, which corresponds to $0.00 \leq\rho\leq 0.67$; medium within-subject correlation, $\sigma^2=0.01$ and $\sigma_0^2=0.04$, which corresponds to $0.50\leq\rho\leq 0.83$; and strong within-subject correlation, $\sigma^2=0.01$ and $\sigma_0^2=0.09$, which corresponds to $0.73\leq\rho\leq 0.91$.

\begin{figure}[!t]
	\centering
	\renewcommand\arraystretch{0}
	\setlength{\tabcolsep}{0pt}
	\begin{tabular}{cccc}
		& \;\;\;\;\; $n=25$ & \;\;\;\;\ $n=50$ & \;\;\;\; $n=100$ \\
		\vspace{-0.5cm}
		\begin{sideways} \hspace{43pt} Weak corr. \end{sideways} &
		\includegraphics[width=0.33\linewidth]{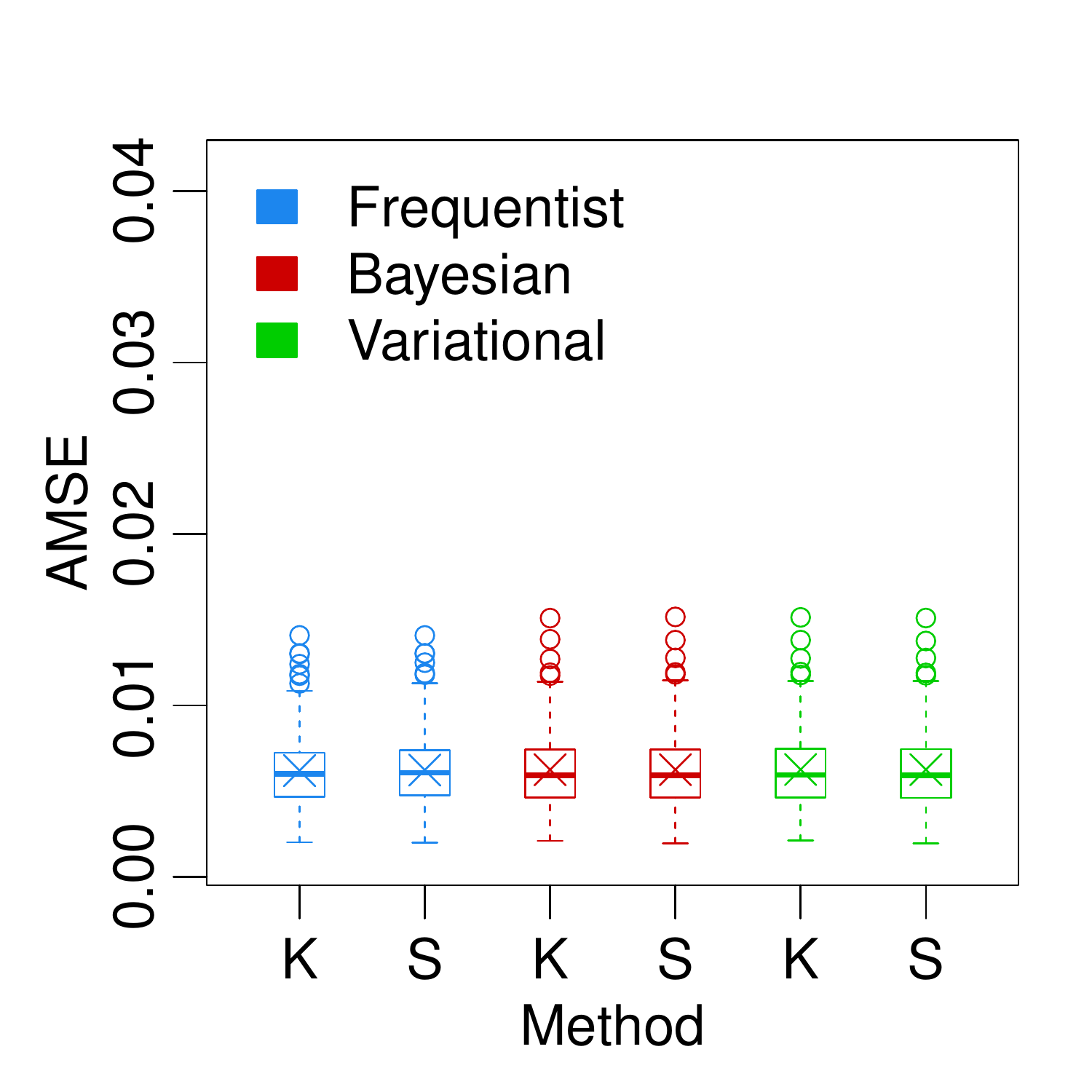}  &
		\includegraphics[width=0.33\linewidth]{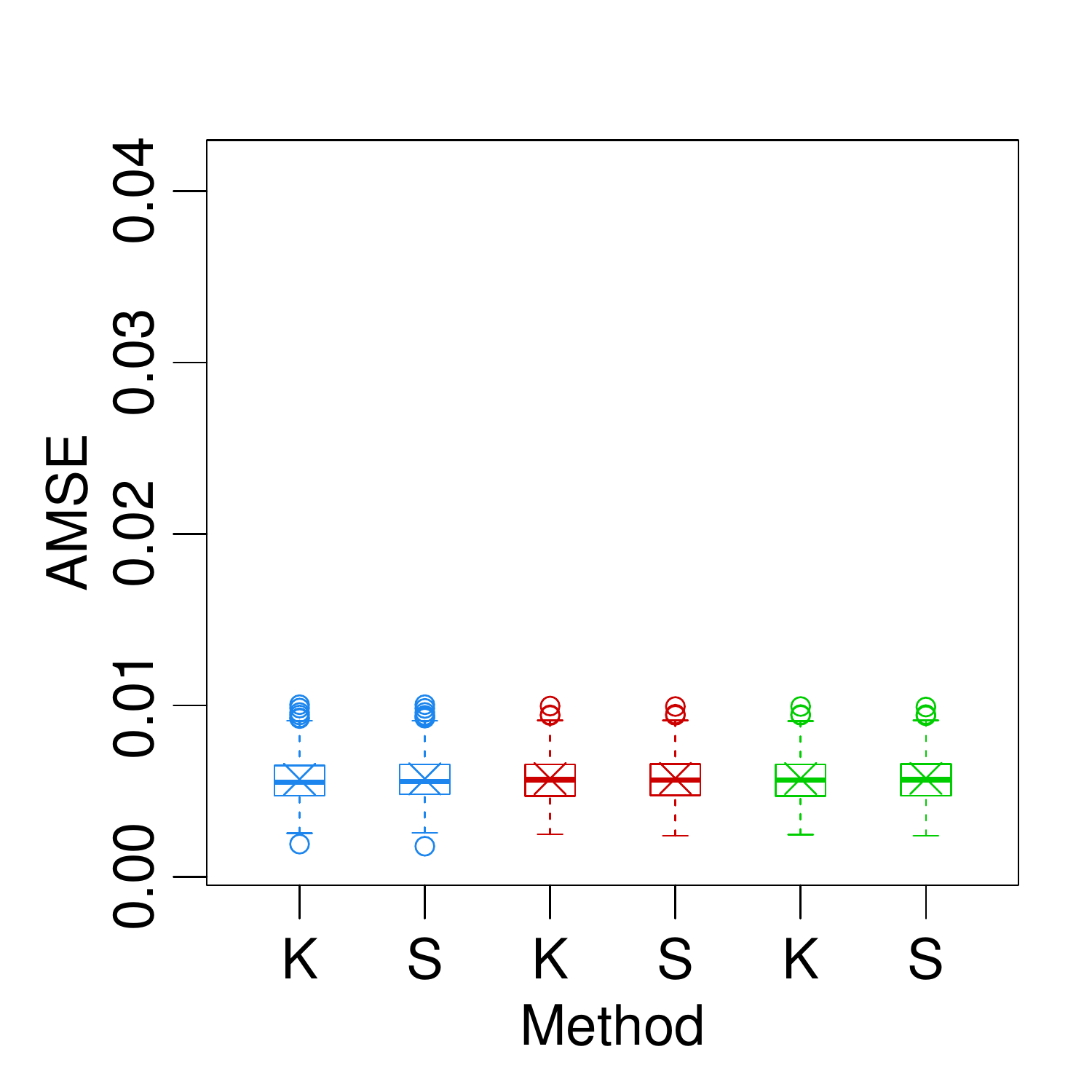}  &
		\includegraphics[width=0.33\linewidth]{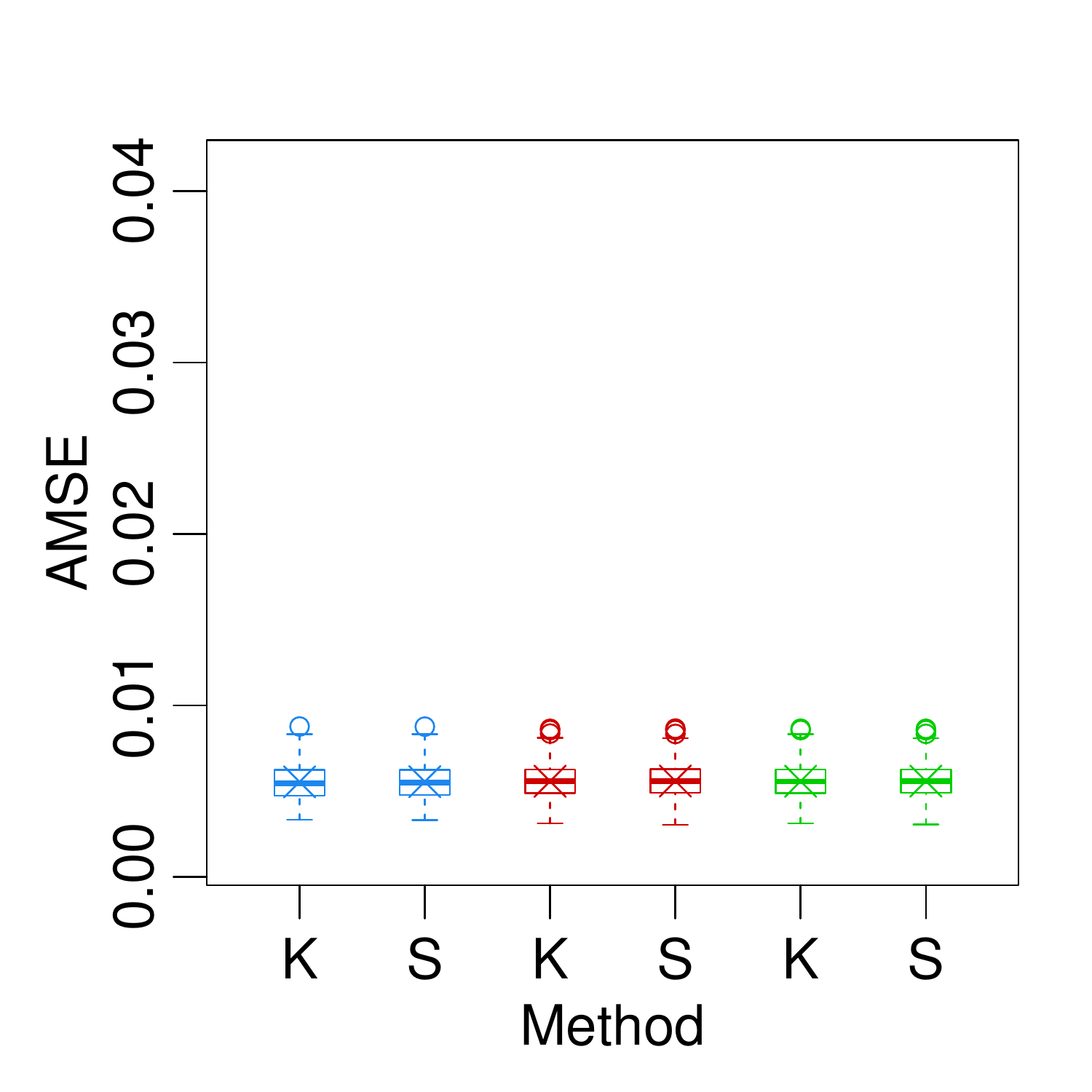} \\
		\vspace{-0.5cm}
		\begin{sideways} \hspace{33pt} Medium corr. \end{sideways}   &
		\includegraphics[width=0.33\linewidth]{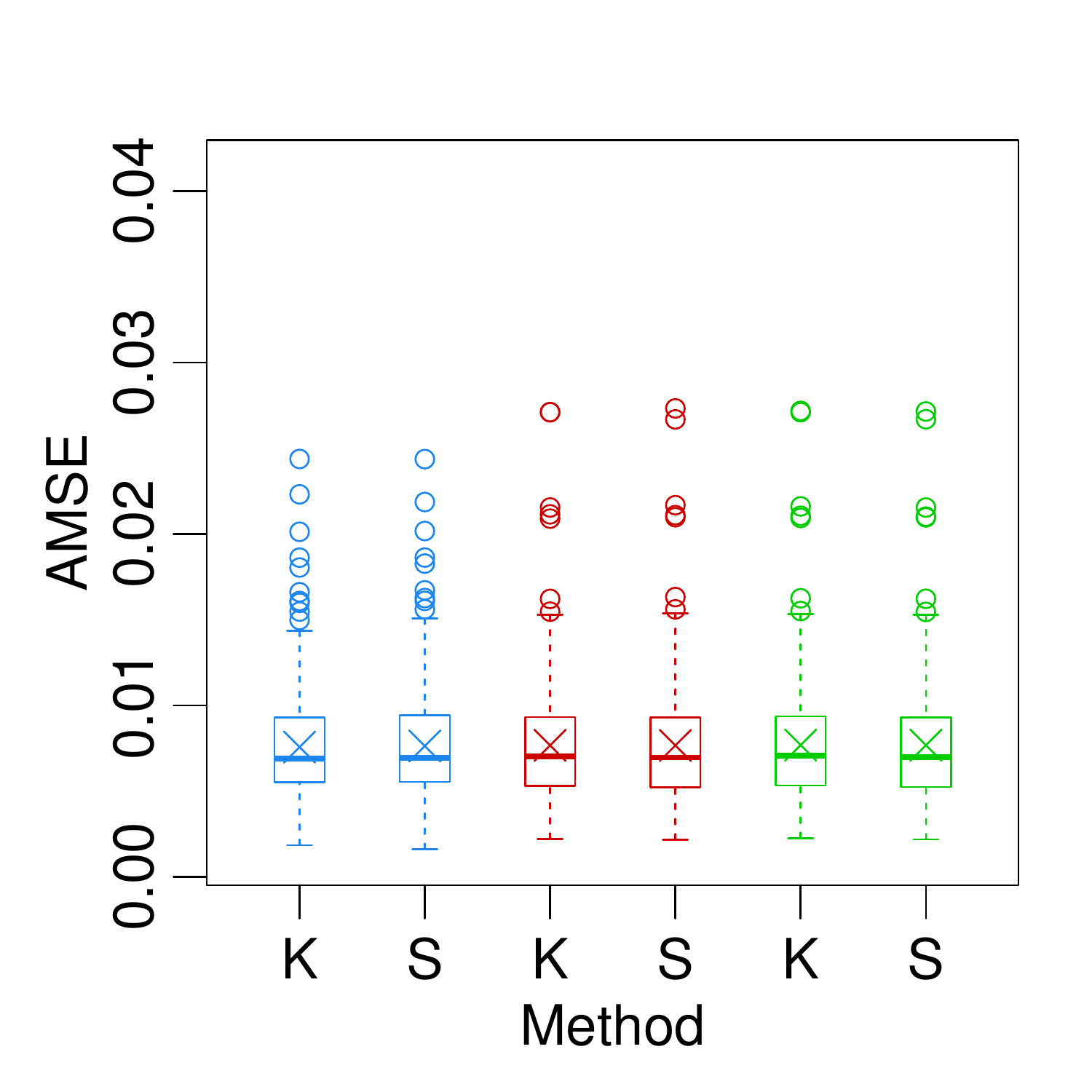} &
		\includegraphics[width=0.33\linewidth]{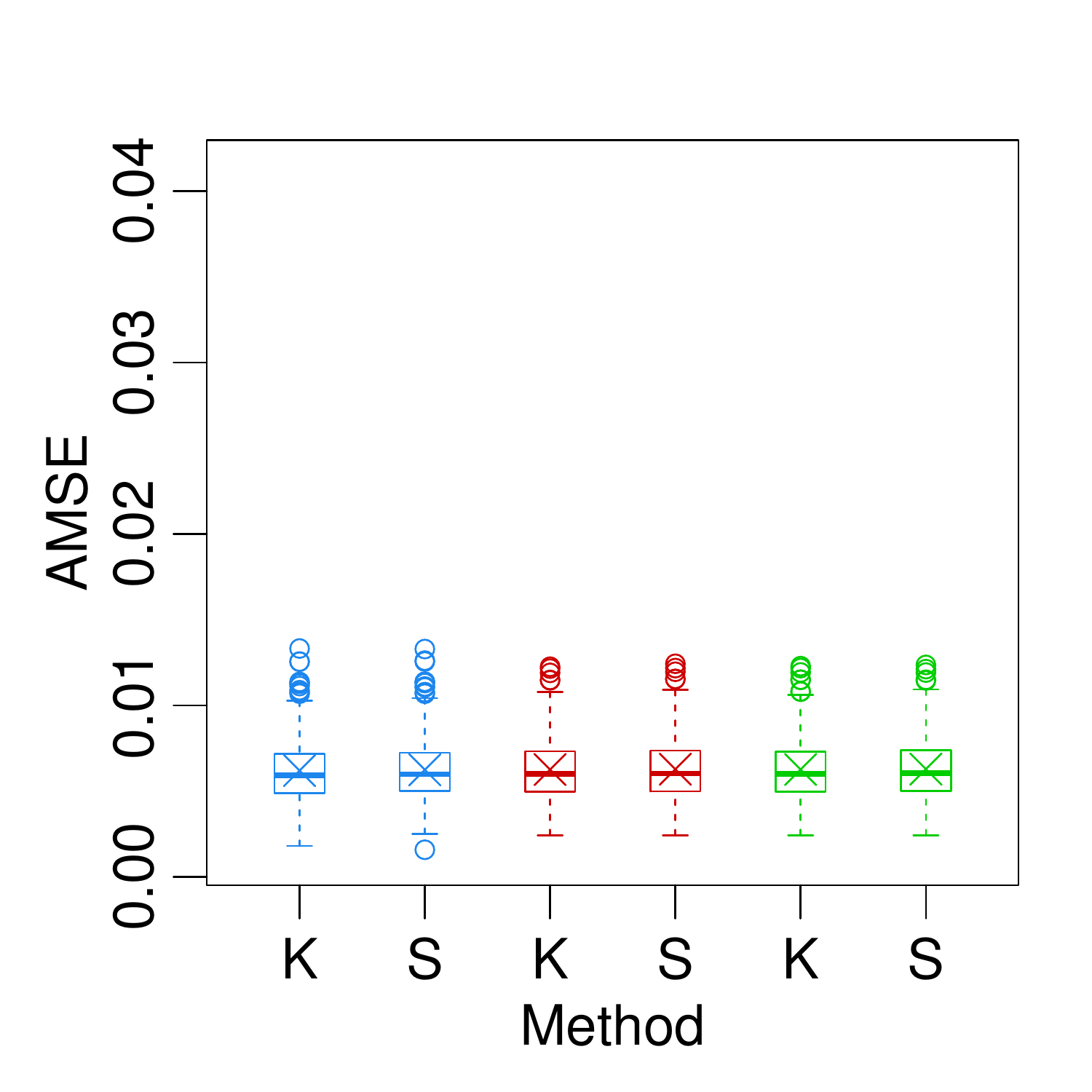} &
		\includegraphics[width=0.33\linewidth]{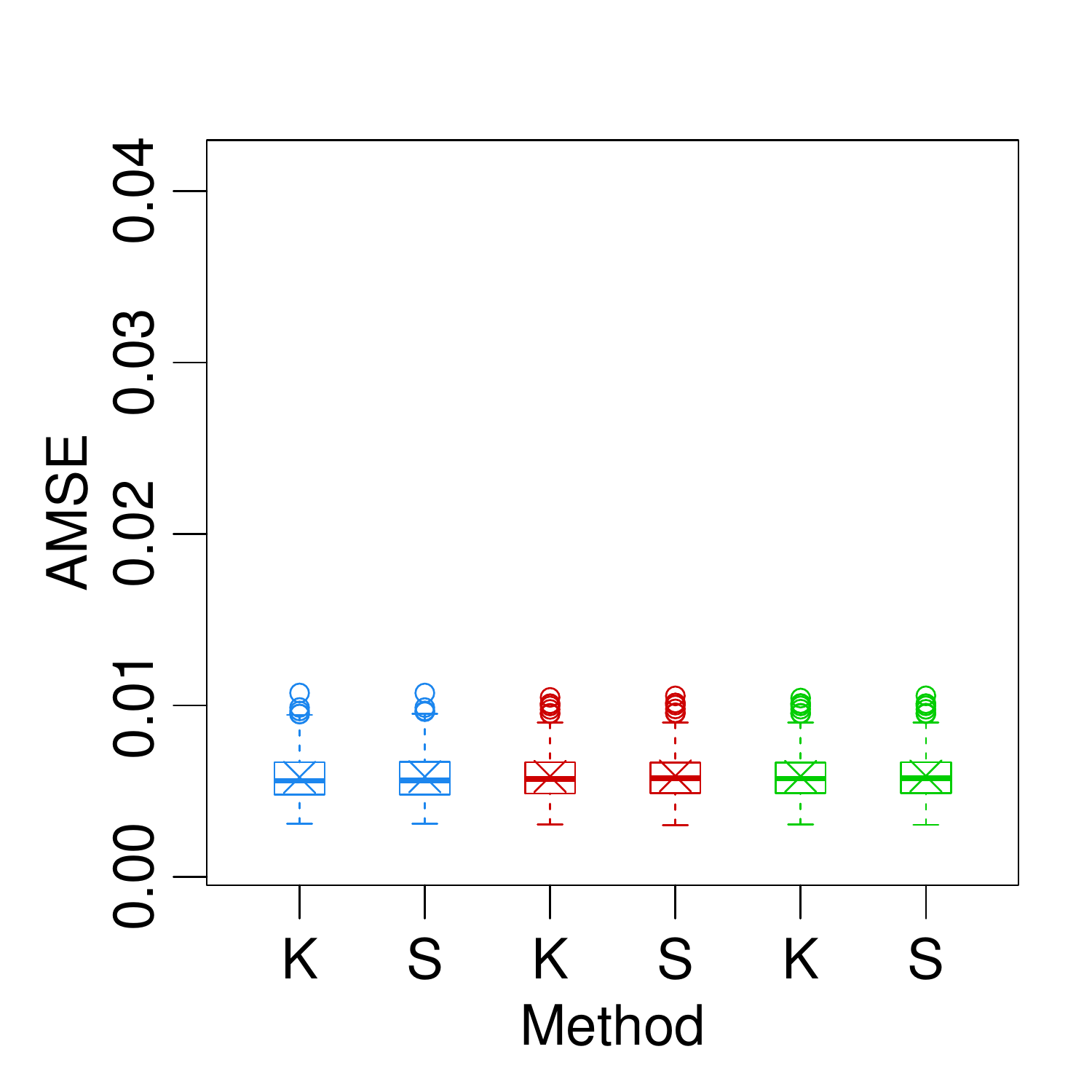} \\
		\begin{sideways} \hspace{43pt} High corr. \end{sideways}   &
		\includegraphics[width=0.33\linewidth]{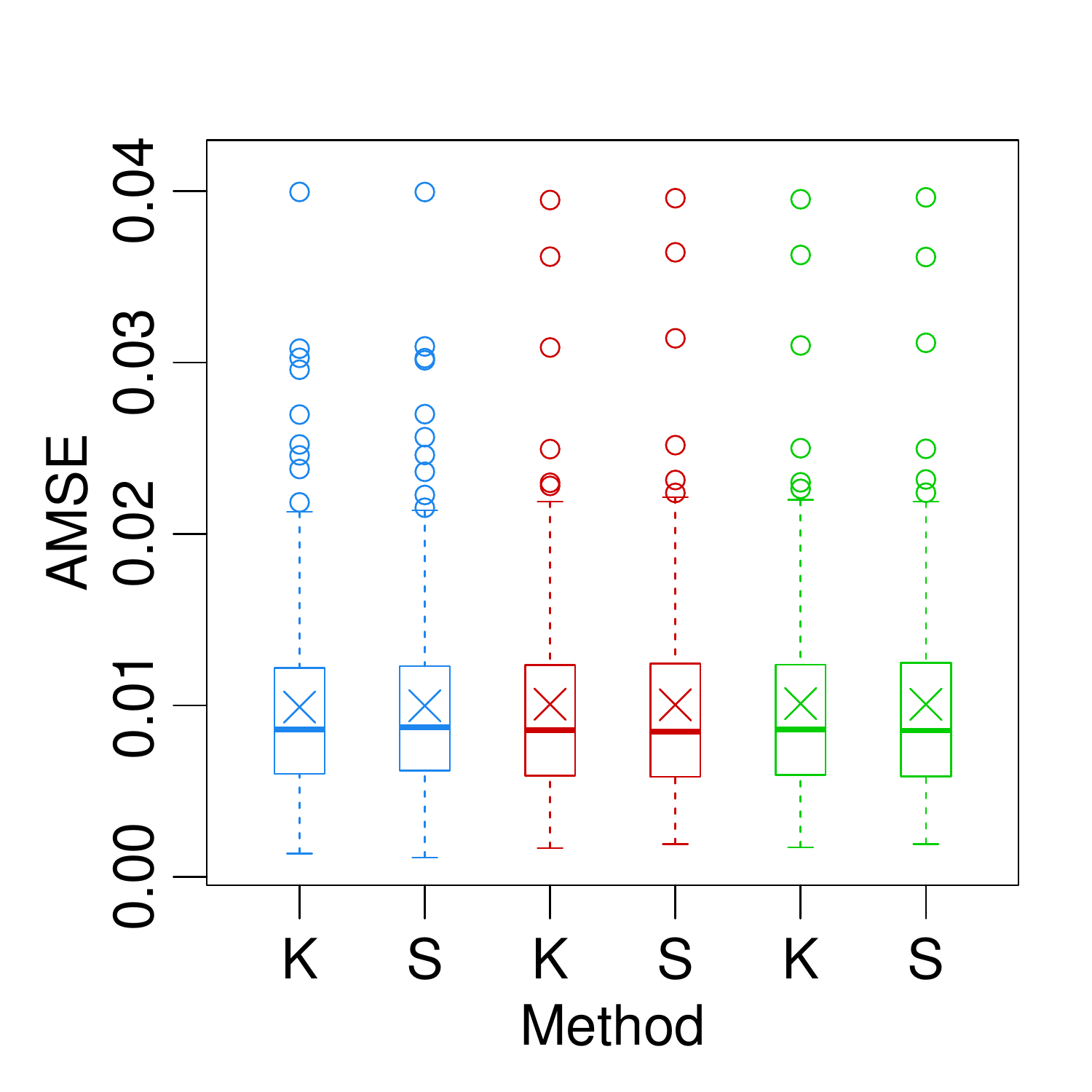} &
		\includegraphics[width=0.33\linewidth]{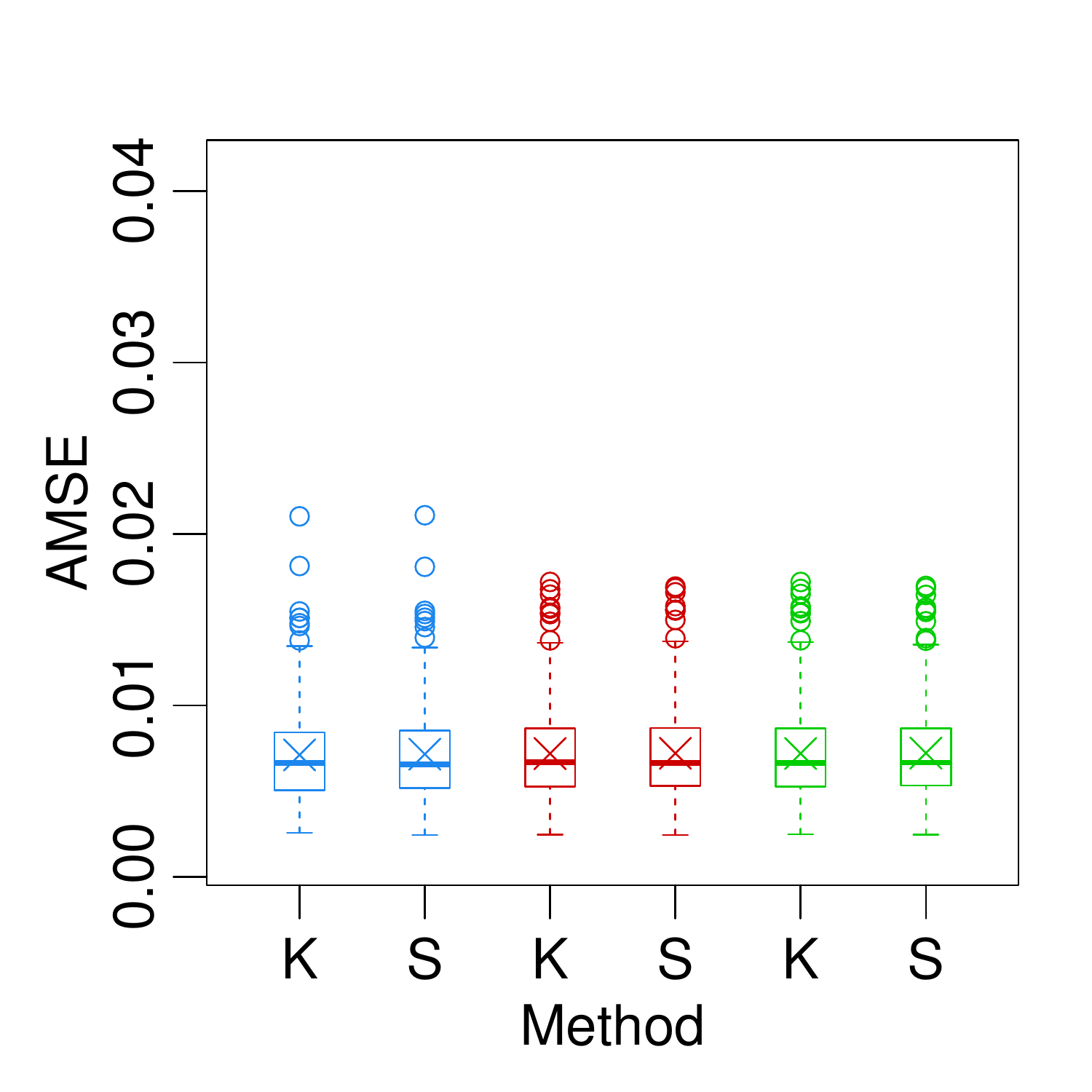} &
		\includegraphics[width=0.33\linewidth]{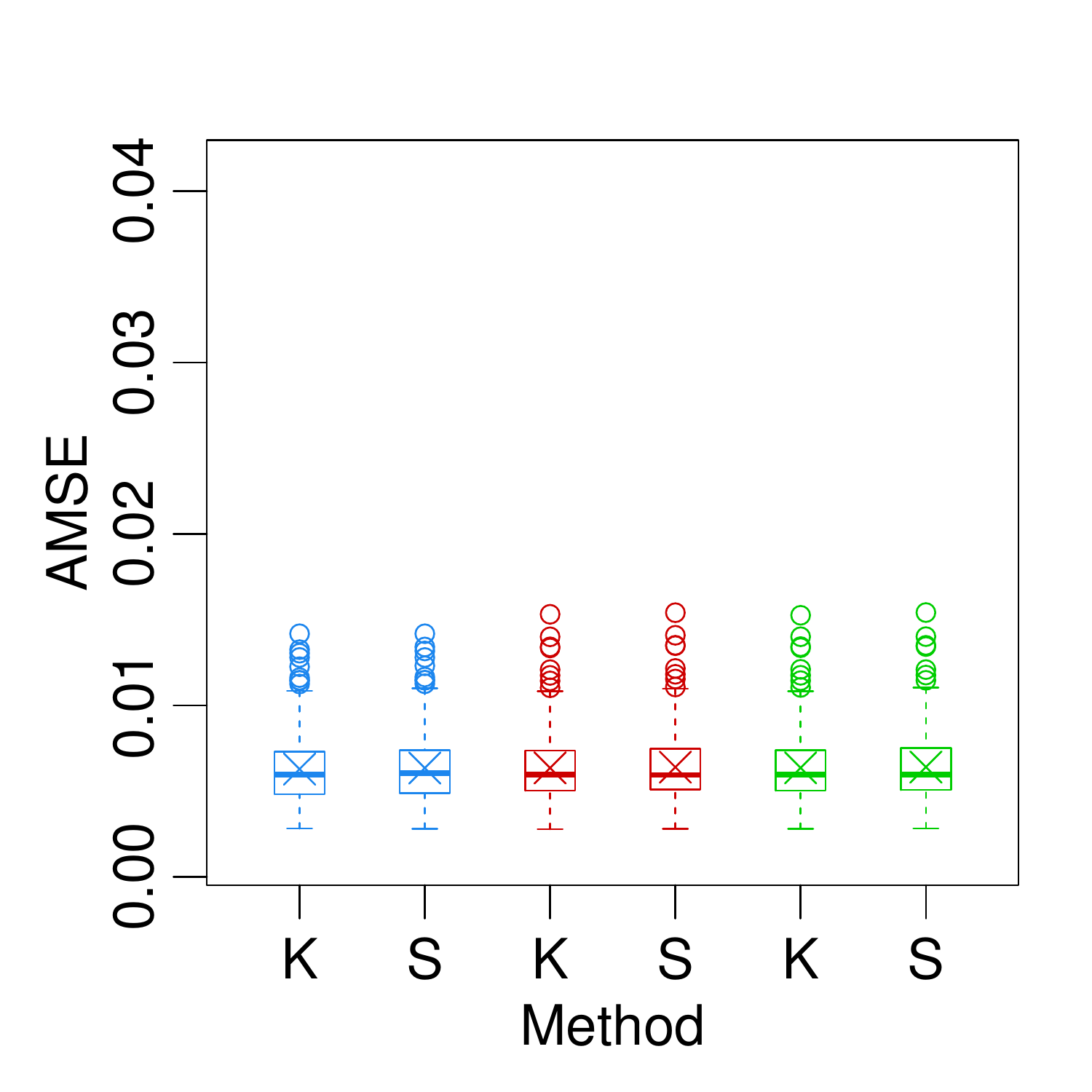} \\
	\end{tabular}
	\caption{\textsf{AMSE} distribution of $\beta_0 = 2\,e^t$ corresponding to 250 synthetic datasets generated according to \textsf{TVCM} \eqref{eq_tvcm_sim1}. Scenarios are delimited by correlation level in rows (weak, medium, and high within-subject correlation) and sample size in columns ($n=25$, $n = 50$, and $n=100$). The model is fitted each time using both radial kernel (\textsf{K}) and regression spline (\textsf{S}) functions, with Frequentist (blue), Bayesian (black), and variational (green) methods.}
	\label{fig_sim1}
\end{figure}

\begin{figure}[!t]
	\centering
	\renewcommand\arraystretch{0}
	\setlength{\tabcolsep}{0pt}
	\begin{tabular}{cccc}
		& \;\;\;\;\; $n=25$ & \;\;\;\;\ $n=50$ & \;\;\;\; $n=100$ \\
		\vspace{-0.5cm}
		\begin{sideways} \hspace{43pt} Weak corr. \end{sideways} &
		\includegraphics[width=0.33\linewidth]{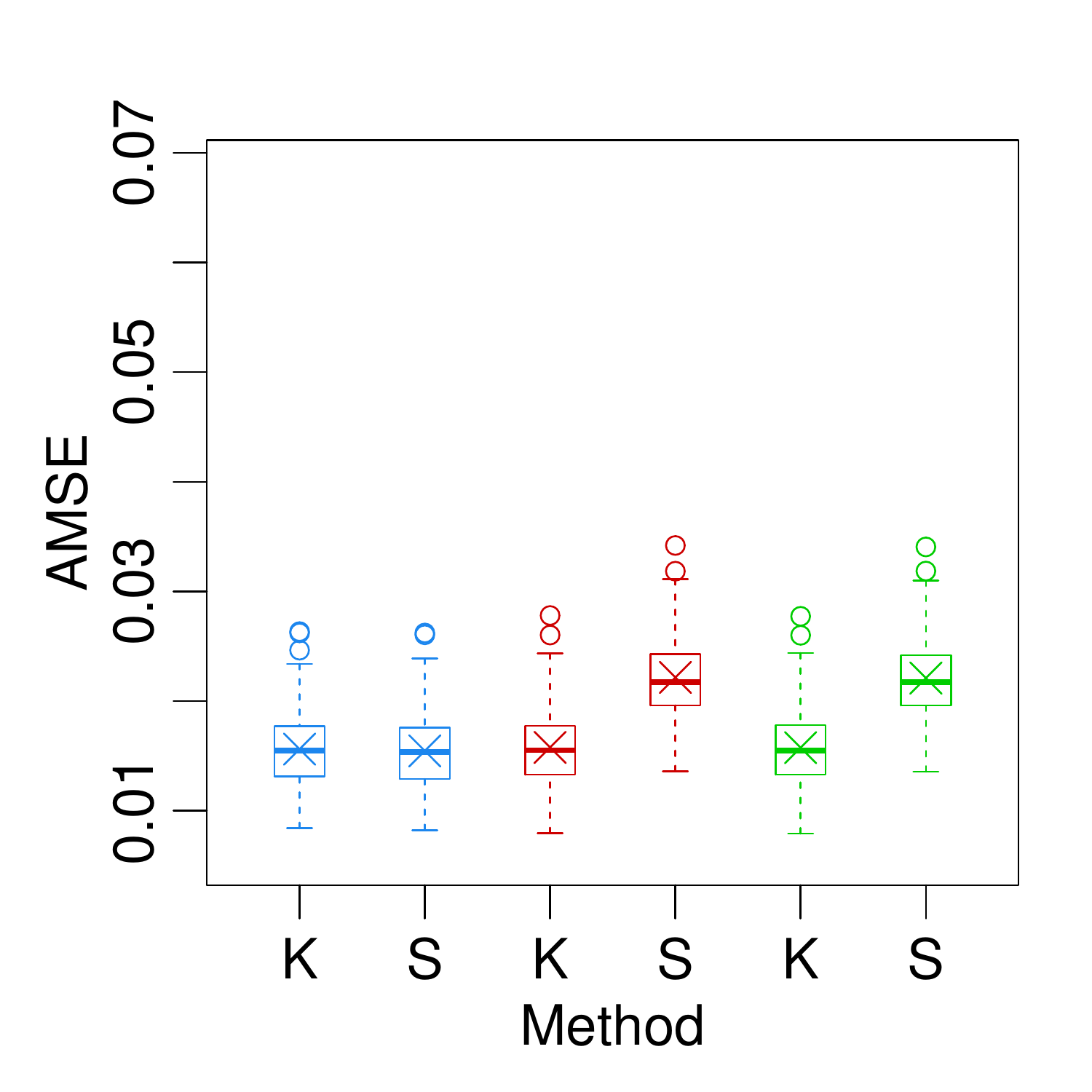}  &
		\includegraphics[width=0.33\linewidth]{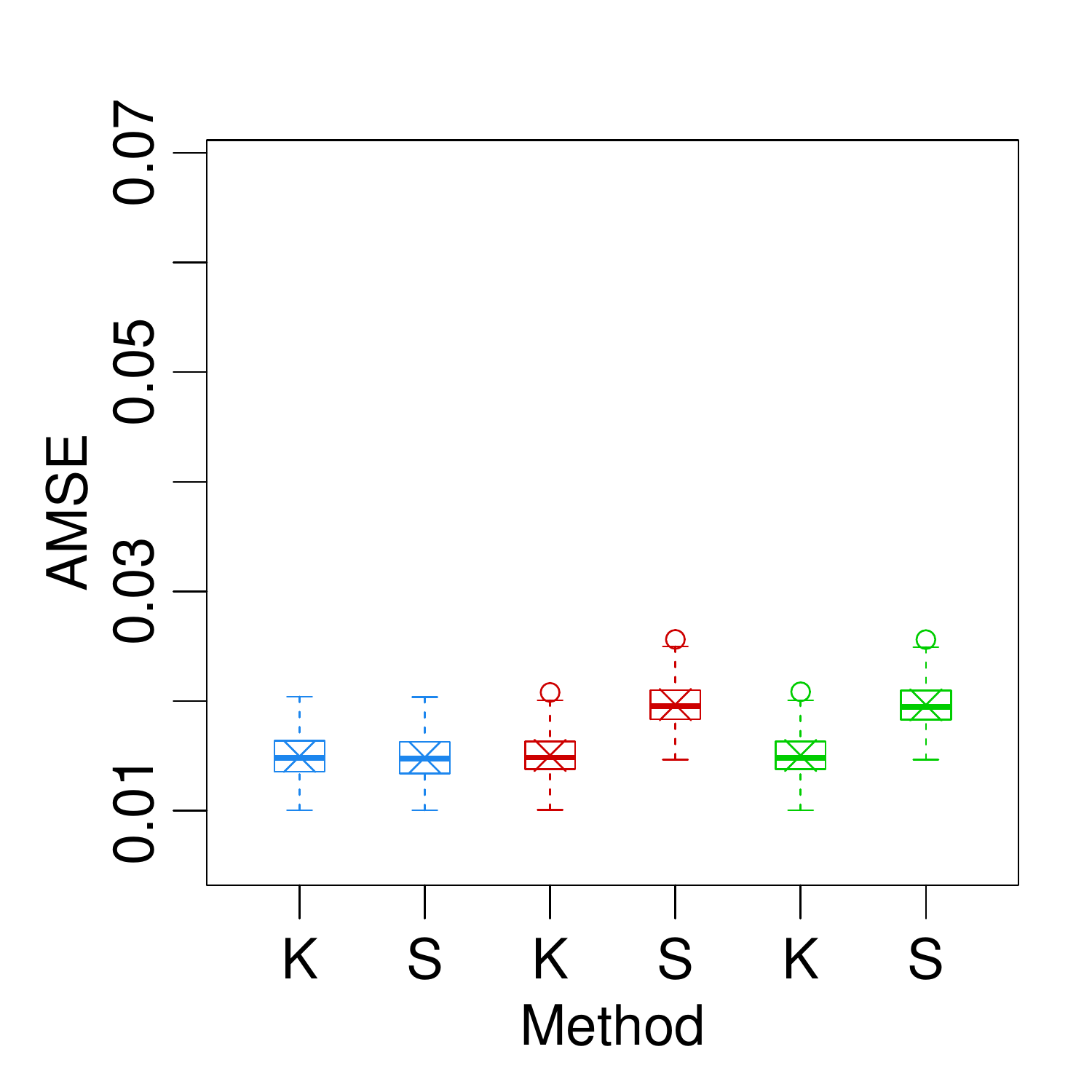}  &
		\includegraphics[width=0.33\linewidth]{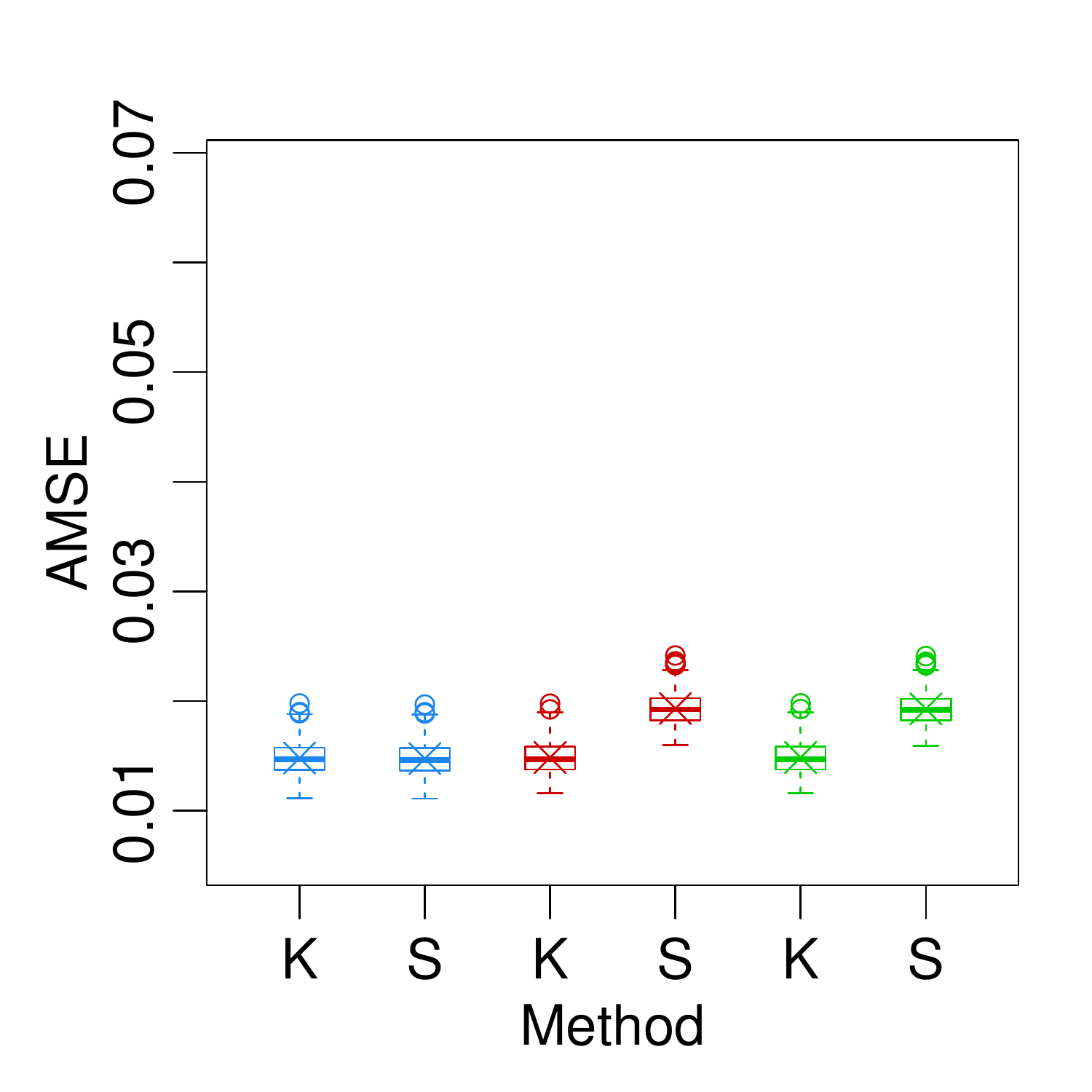} \\
		\vspace{-0.5cm}
		\begin{sideways} \hspace{33pt} Medium corr. \end{sideways}   &
		\includegraphics[width=0.33\linewidth]{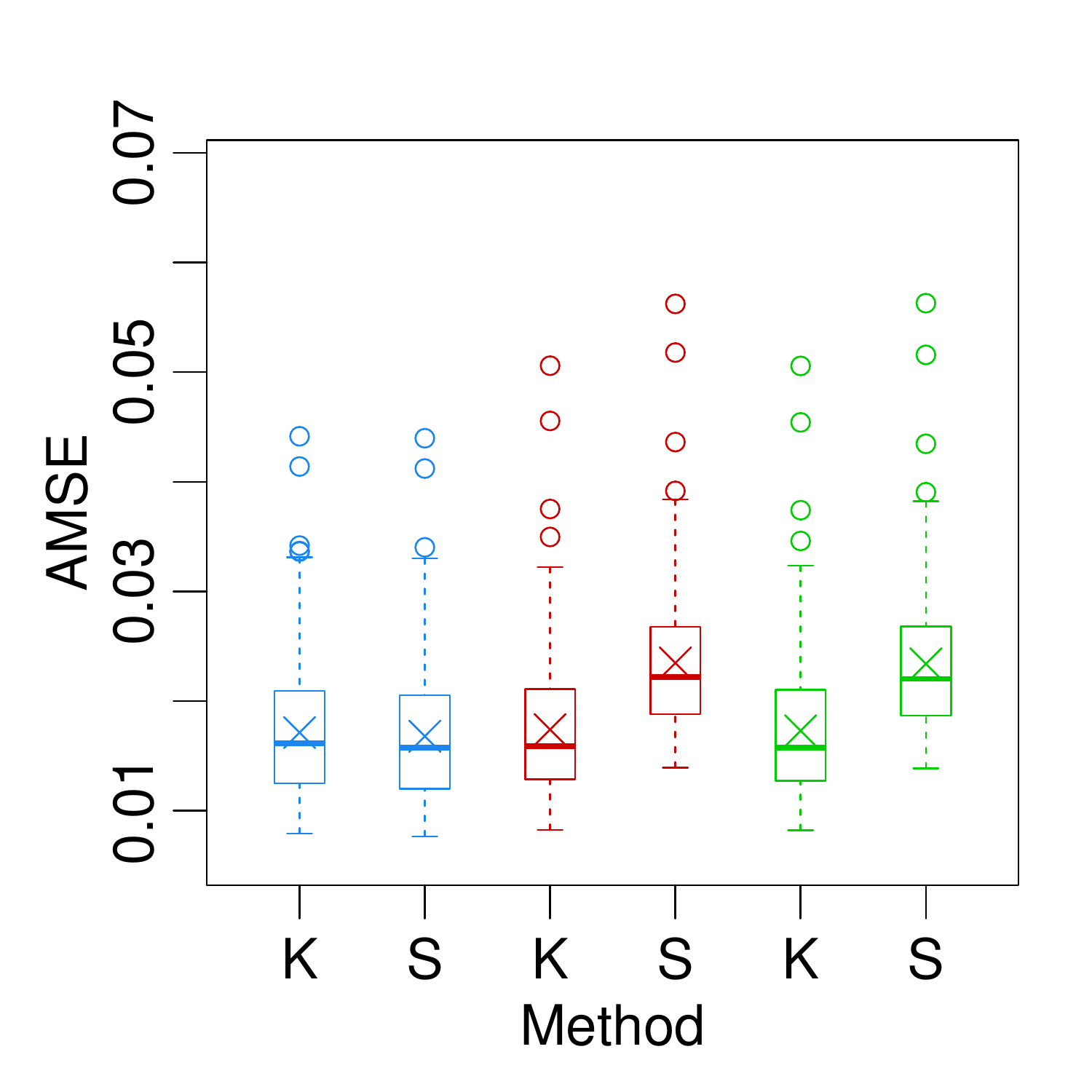} &
		\includegraphics[width=0.33\linewidth]{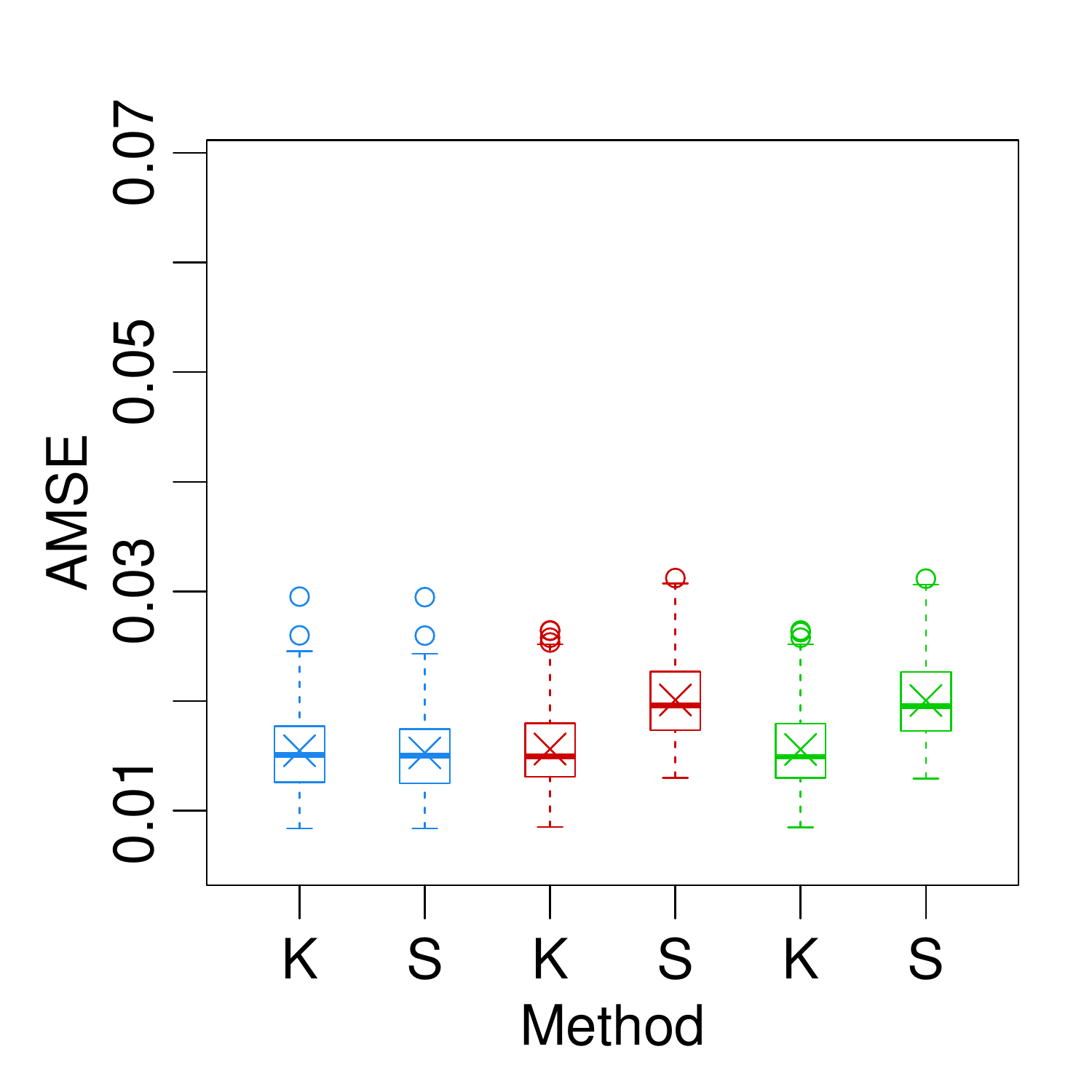} &
		\includegraphics[width=0.33\linewidth]{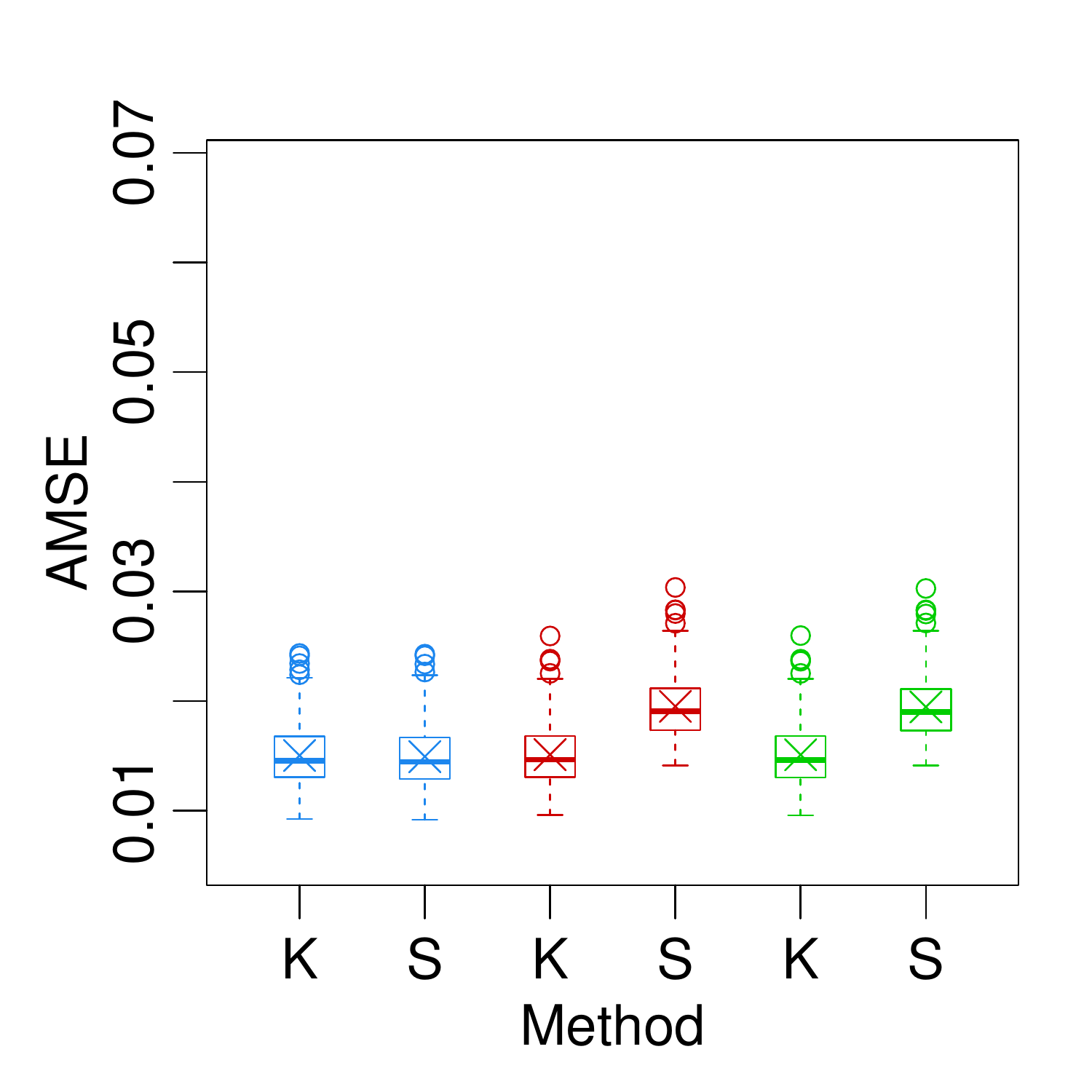} \\
		\begin{sideways} \hspace{43pt} High corr. \end{sideways}   &
		\includegraphics[width=0.33\linewidth]{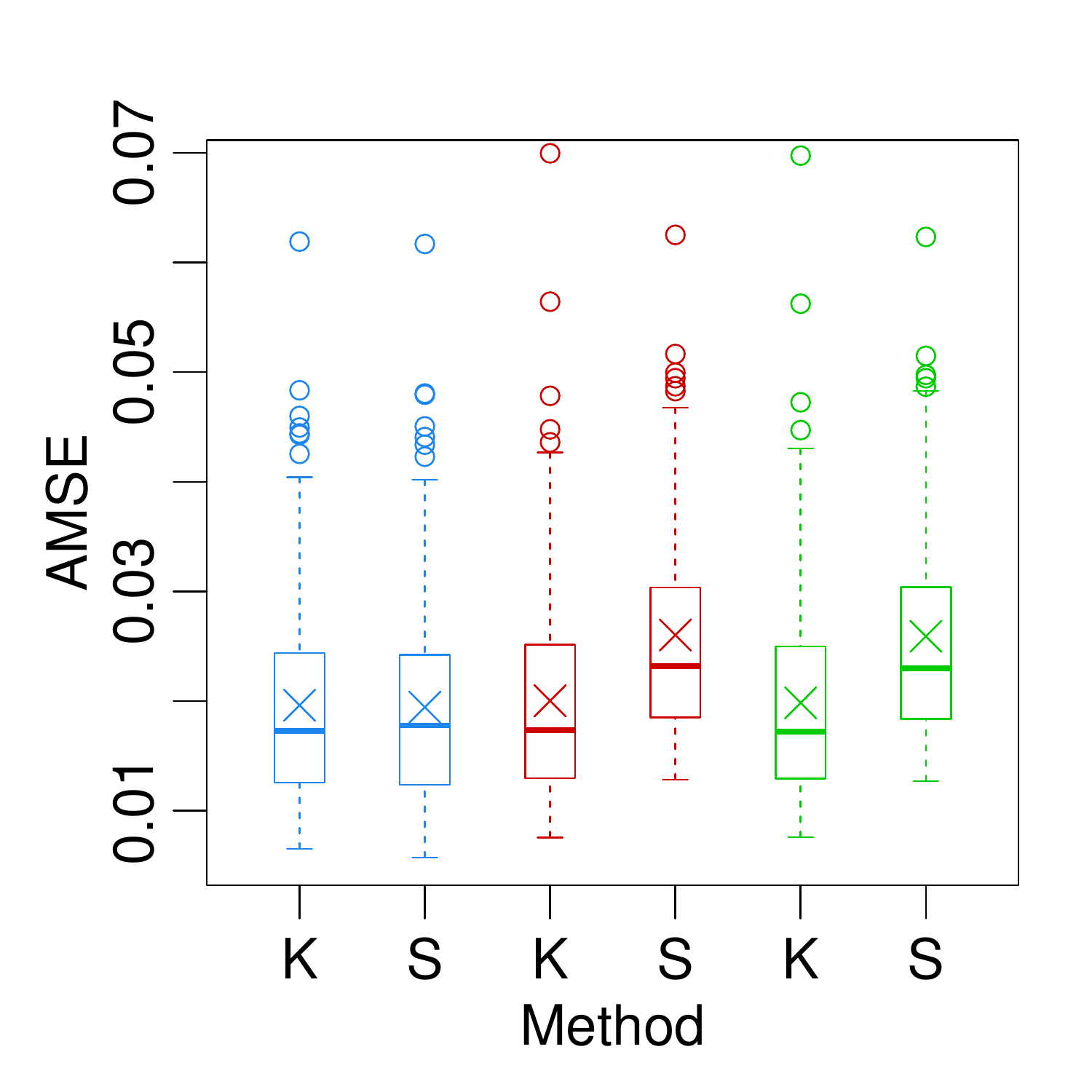} &
		\includegraphics[width=0.33\linewidth]{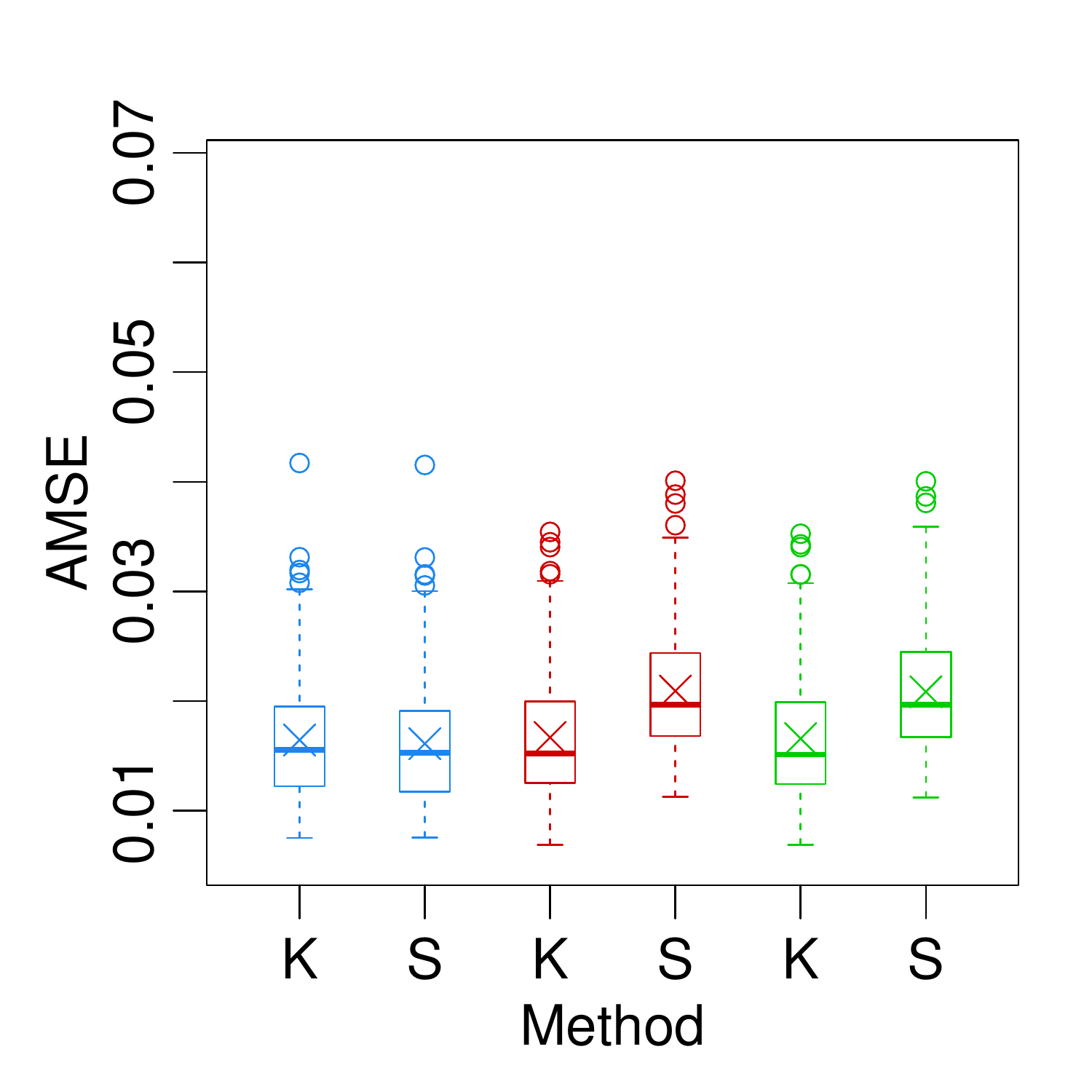} &
		\includegraphics[width=0.33\linewidth]{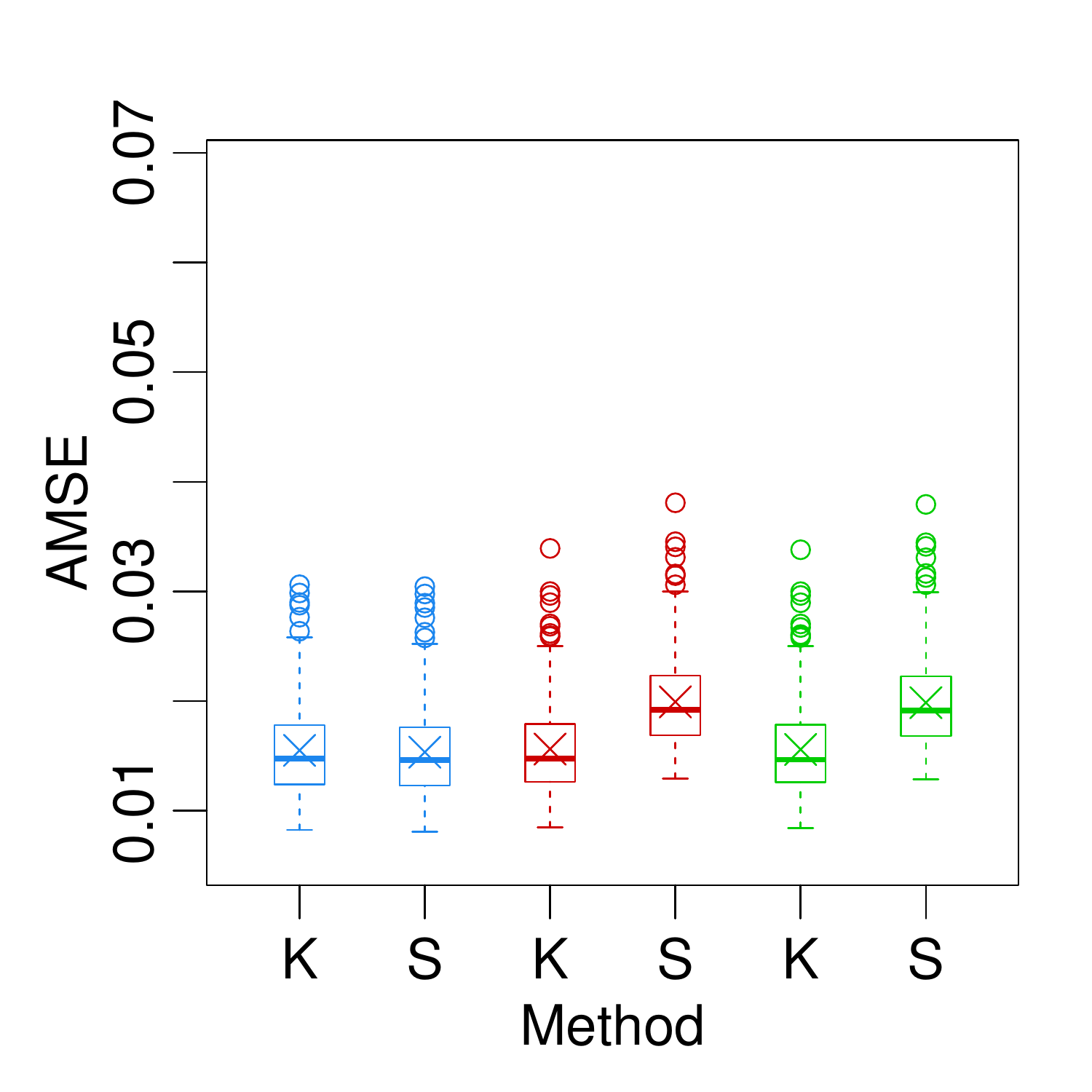} \\
	\end{tabular}
	\caption{\textsf{AMSE} distribution of $\beta_0(t) =  1 + \cos(2 \pi t) + \sin(2 \pi t)$ corresponding to 250 synthetic datasets generated according to \textsf{TVCM} \eqref{eq_tvcm_sim1}. Scenarios are delimited by correlation level in rows (weak, medium, and high within-subject correlation) and sample size in columns ($n=25$, $n = 50$, and $n=100$). The model is fitted each time using both radial kernel (\textsf{K}) and regression spline (\textsf{S}) functions, with Frequentist (blue), Bayesian (black), and variational (green) methods.}
	\label{fig_sim2}
\end{figure}

Additionally, design times are simulated as $t_{i,j}=j/(m+1)$, $i=1,\ldots,n$, $j=1,\ldots,m$, where $m$ is a positive integer. In order to simulate unbalanced datasets for each subject, repeated measures are randomly removed with a rate $r=0.5$; thus, we expect $m(1-r)$ repeated measurements per experimental unit and $nm(1-r)$ measurements in total. In addition, {\color{black} number an location} of knots are chosen according to the \textsf{PCV} criteria {\color{black} and the equally spaced method} described in Section \ref{sec_selection_number_location_knots}, {\color{black} respectively.} We generated $250$ datasets with two dynamic parameters, $\beta_0(t) = 2e^t$ and $\beta_0(t) =  1 + \cos(2 \pi t) + \sin(2 \pi t)$, and also,  three sample sizes, $n=25$, $n=50$, and $n=100$. Each time, {\color{black} once the number and location of knots are fixed,} we fitted model \eqref{eq_tvcm} using both radial kernel and regression spline functions setting $g=2$ as a degree, with Frequentist, Bayesian, and variational methods. {\color{black} Setting the prior distribution as discussed in Section \ref{sec_bayesian_inference},} Bayesian and variational estimates are based on $2,000$ samples {\color{black} from the posterior distribution. For Bayesian inference, we use a burn-in period of 500 samples; whereas for variational inference, we use a negligible increased in \textsf{ELBO} of \texttt{1e-06}. Such a setting showed no evidence of lack of convergence in any case.}

\begin{figure}[!t]
	\centering
	\renewcommand\arraystretch{0}
	\setlength{\tabcolsep}{0pt}
	\begin{tabular}{cccc}
		& \;\;\;\;\; $n=25$ & \;\;\;\;\ $n=50$ & \;\;\;\; $n=100$ \\
		\vspace{-0.5cm}
		\begin{sideways} \hspace{49pt} \textsf{MADE} \end{sideways} &
		\includegraphics[width=0.33\linewidth]{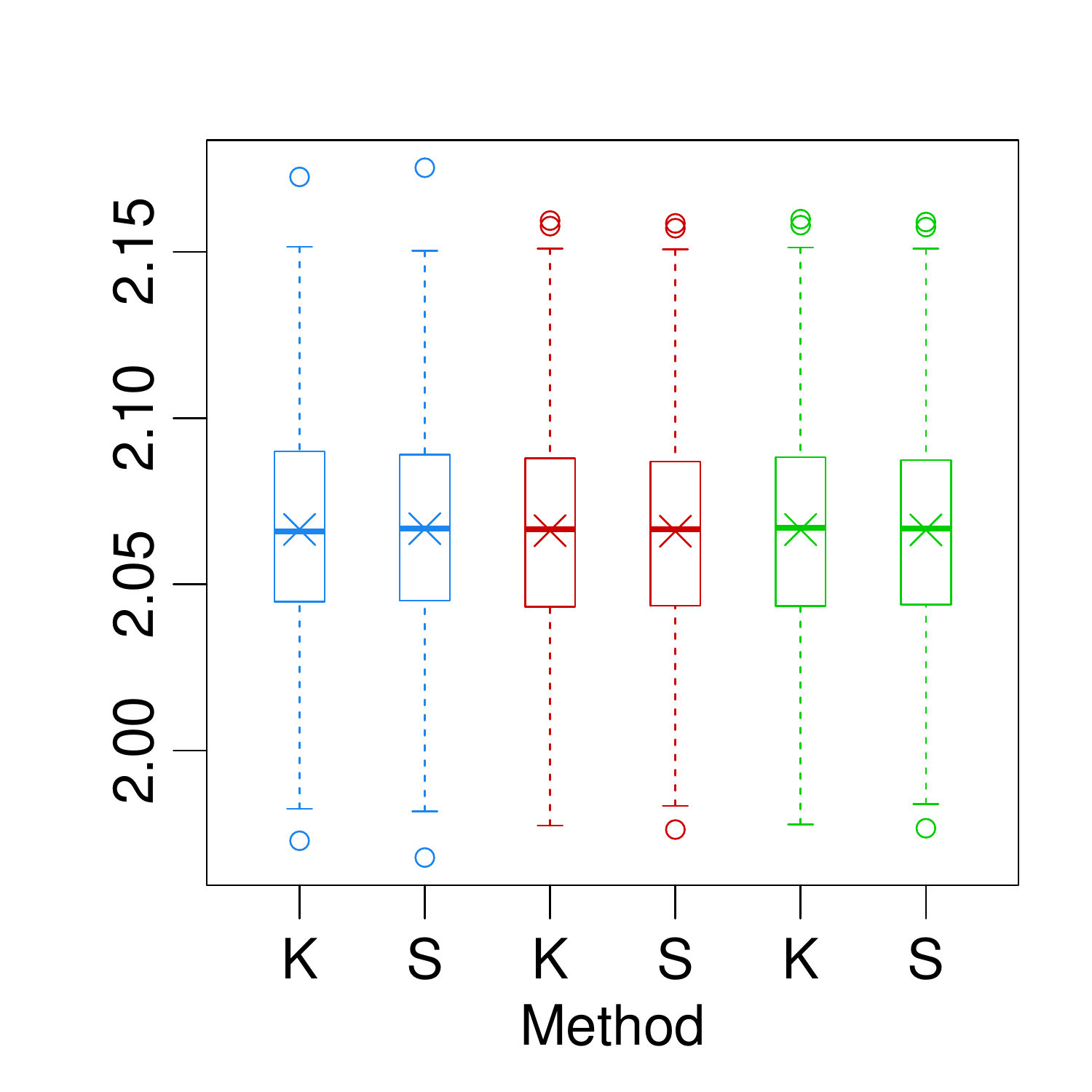}  &
		\includegraphics[width=0.33\linewidth]{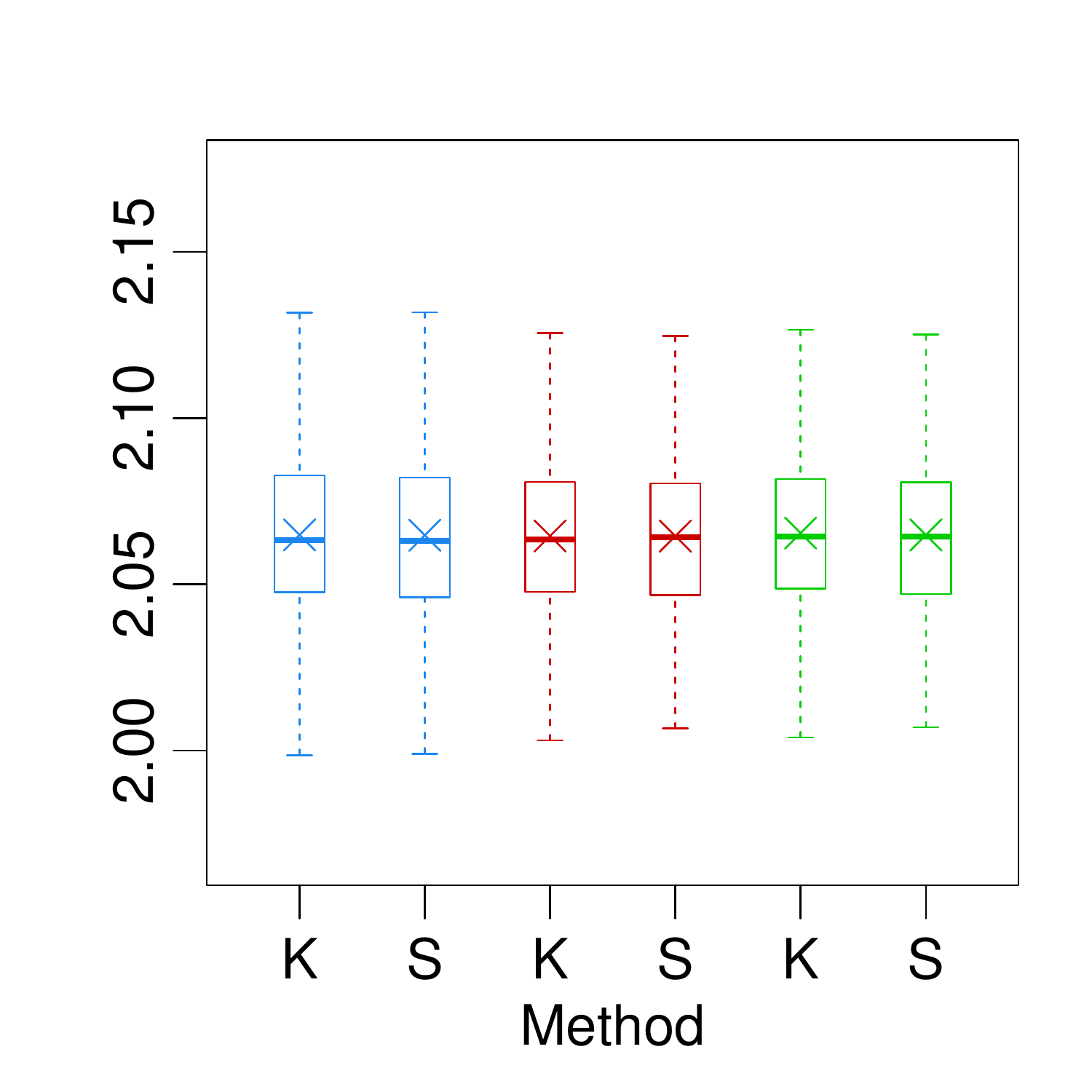}  &
		\includegraphics[width=0.33\linewidth]{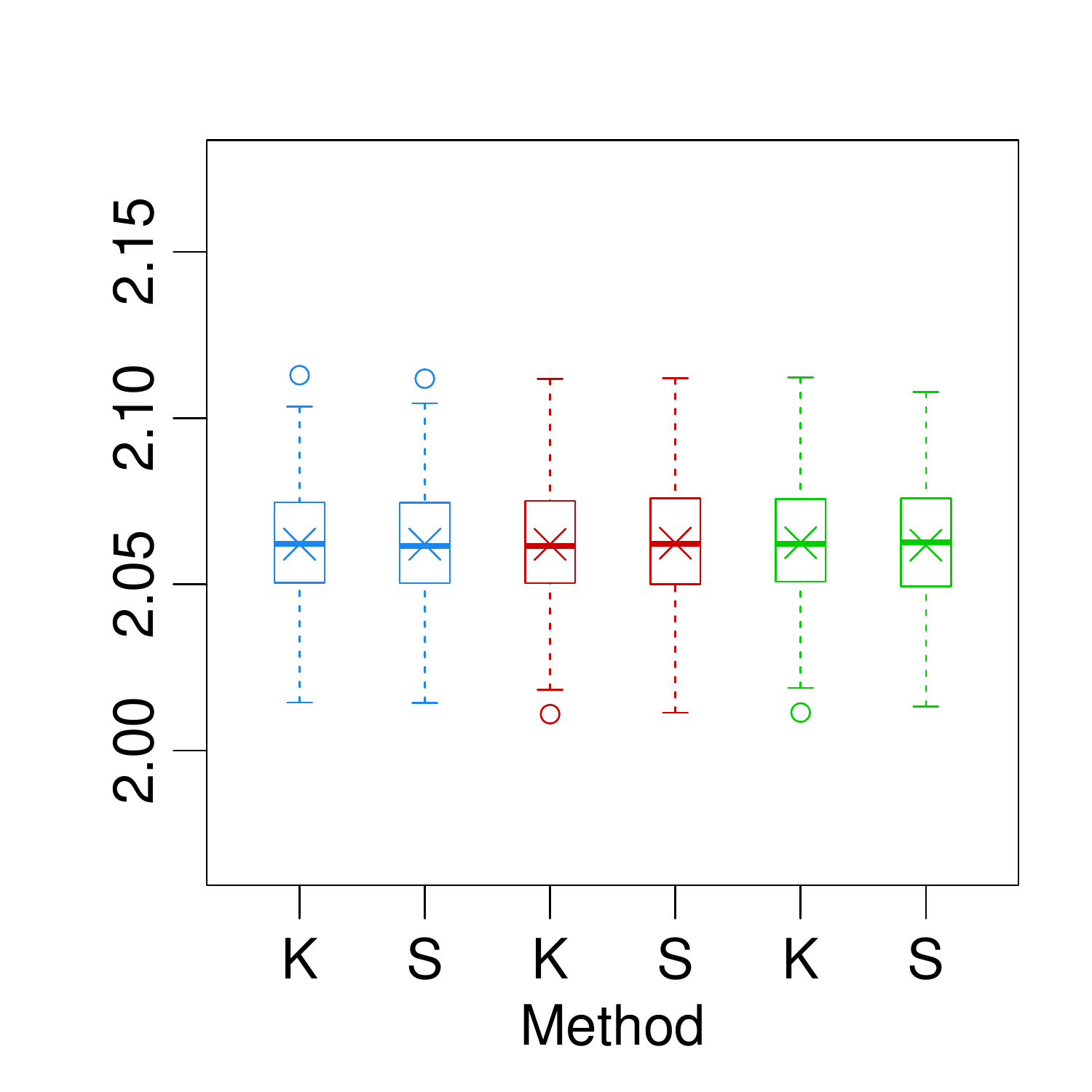} \\
		\vspace{-0.5cm}
		\begin{sideways} \hspace{55pt} $\beta_0(t)$ \end{sideways}   &
		\includegraphics[width=0.33\linewidth]{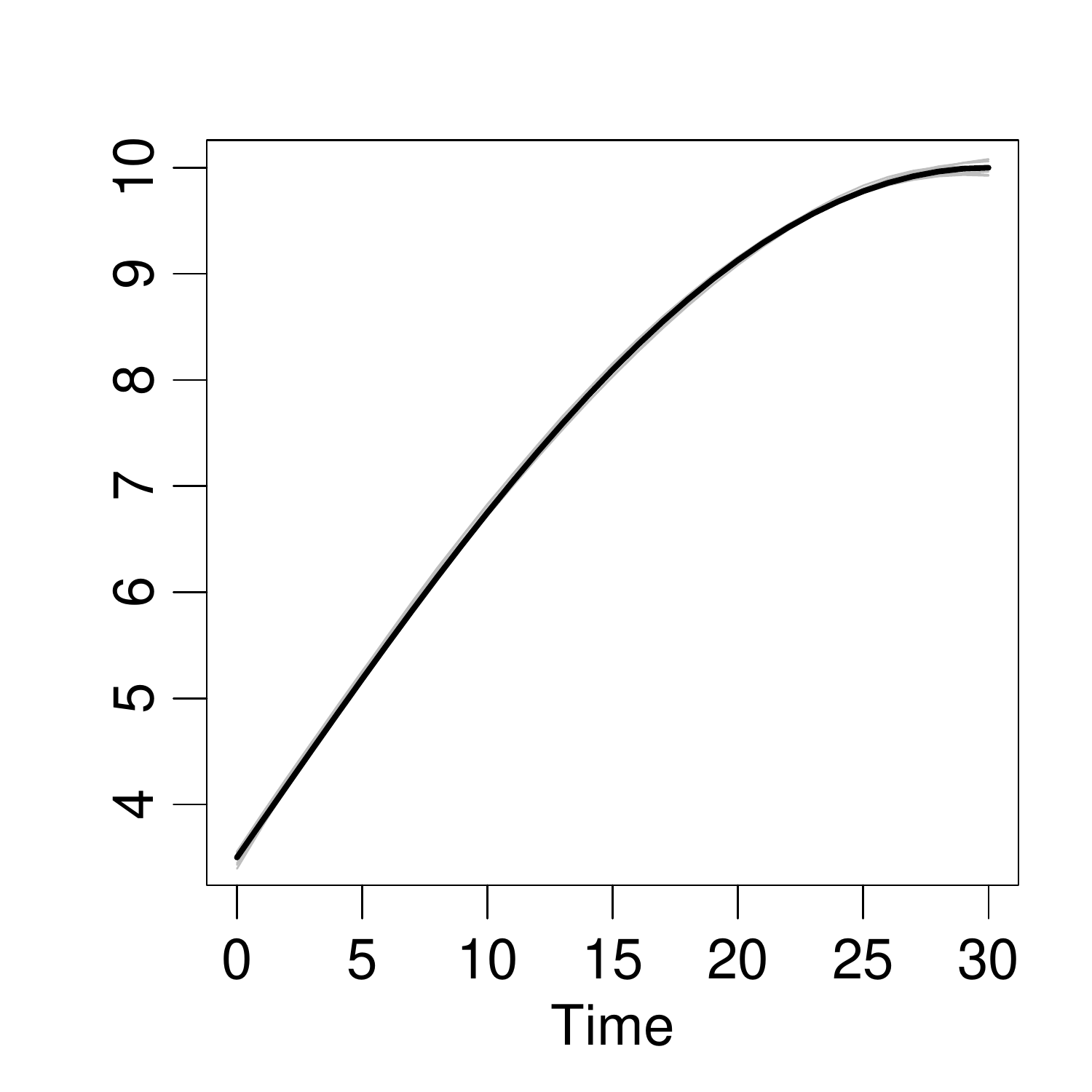} &
		\includegraphics[width=0.33\linewidth]{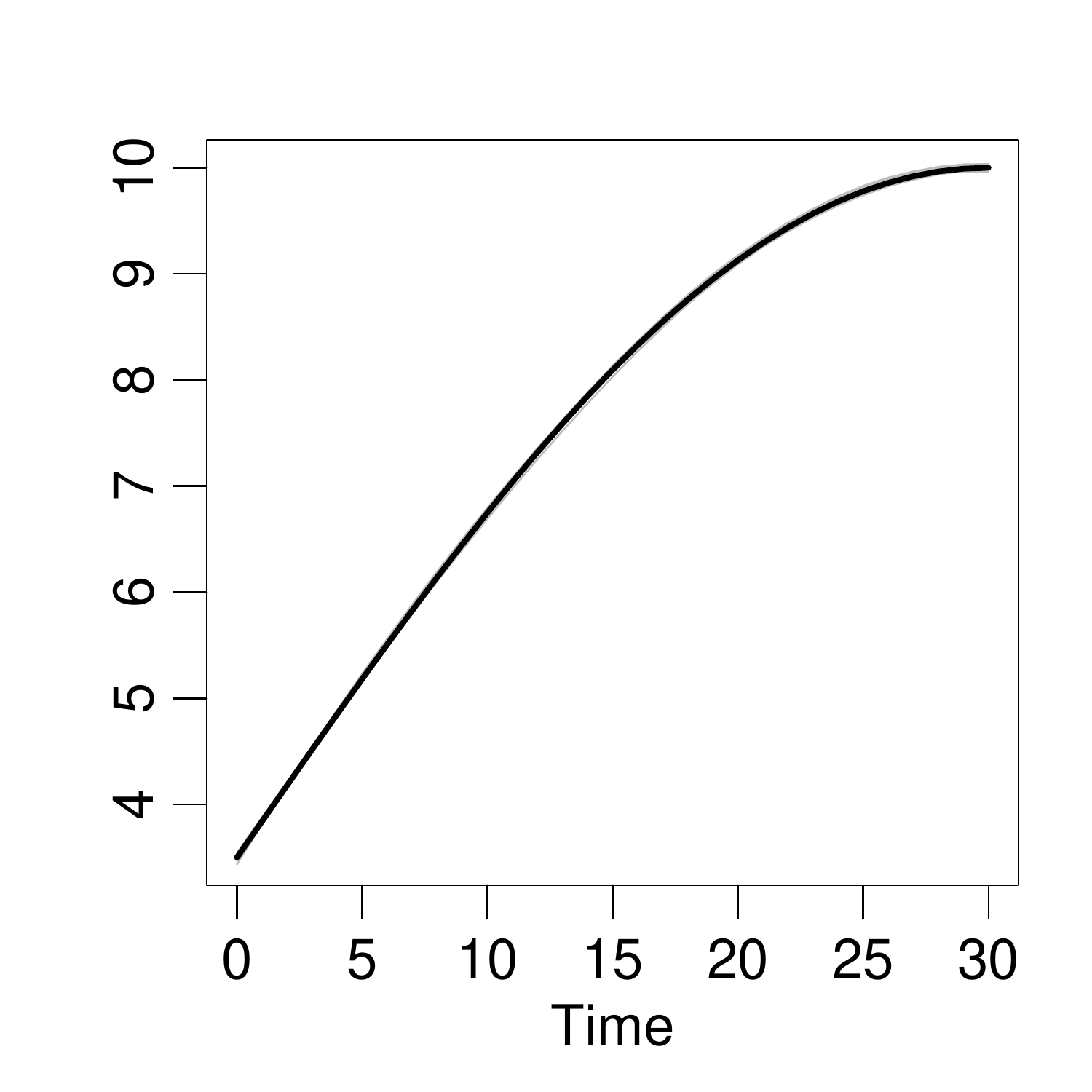} &
		\includegraphics[width=0.33\linewidth]{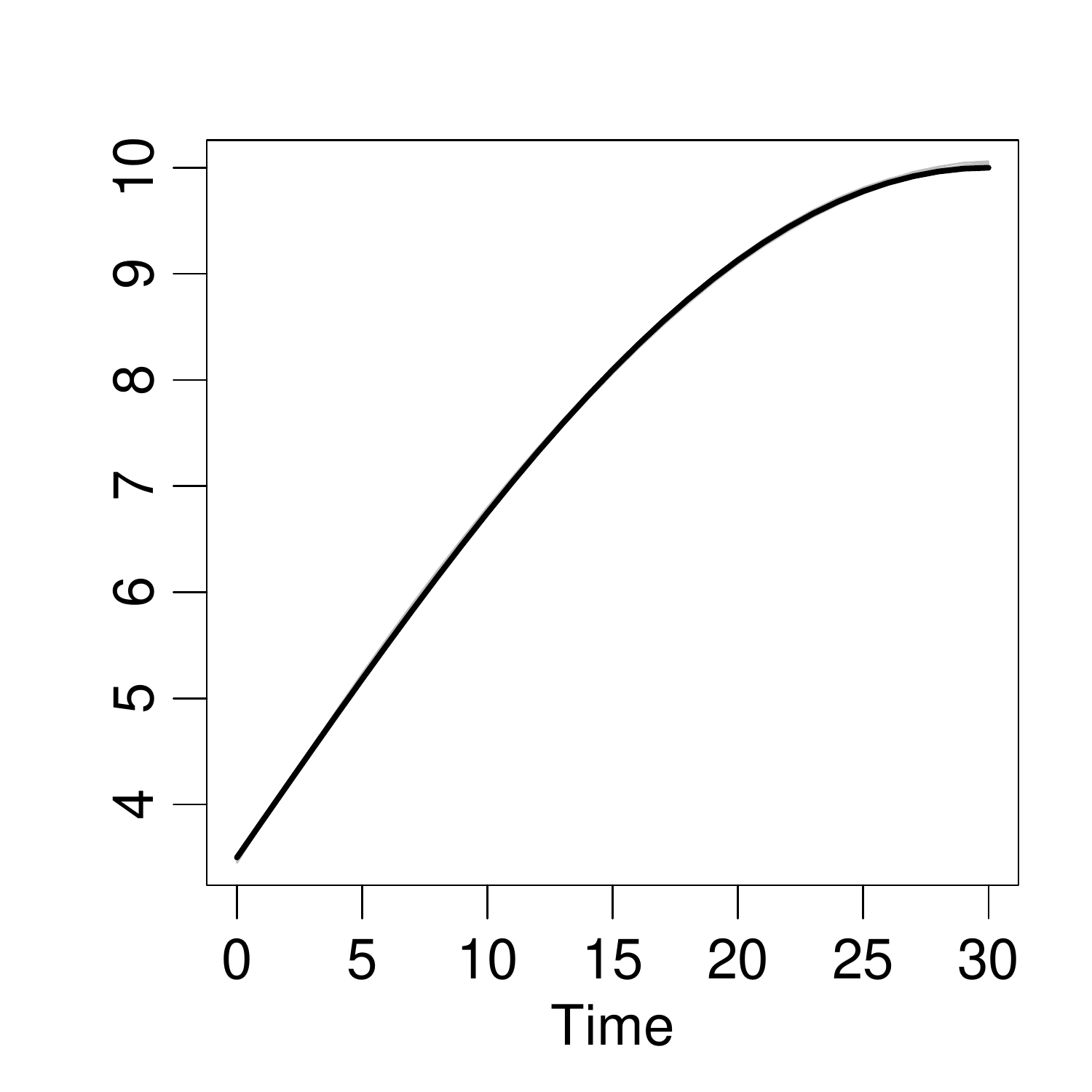} \\
		\vspace{-0.5cm}
		\begin{sideways} \hspace{55pt} $\beta_1(t)$ \end{sideways}   &
		\includegraphics[width=0.33\linewidth]{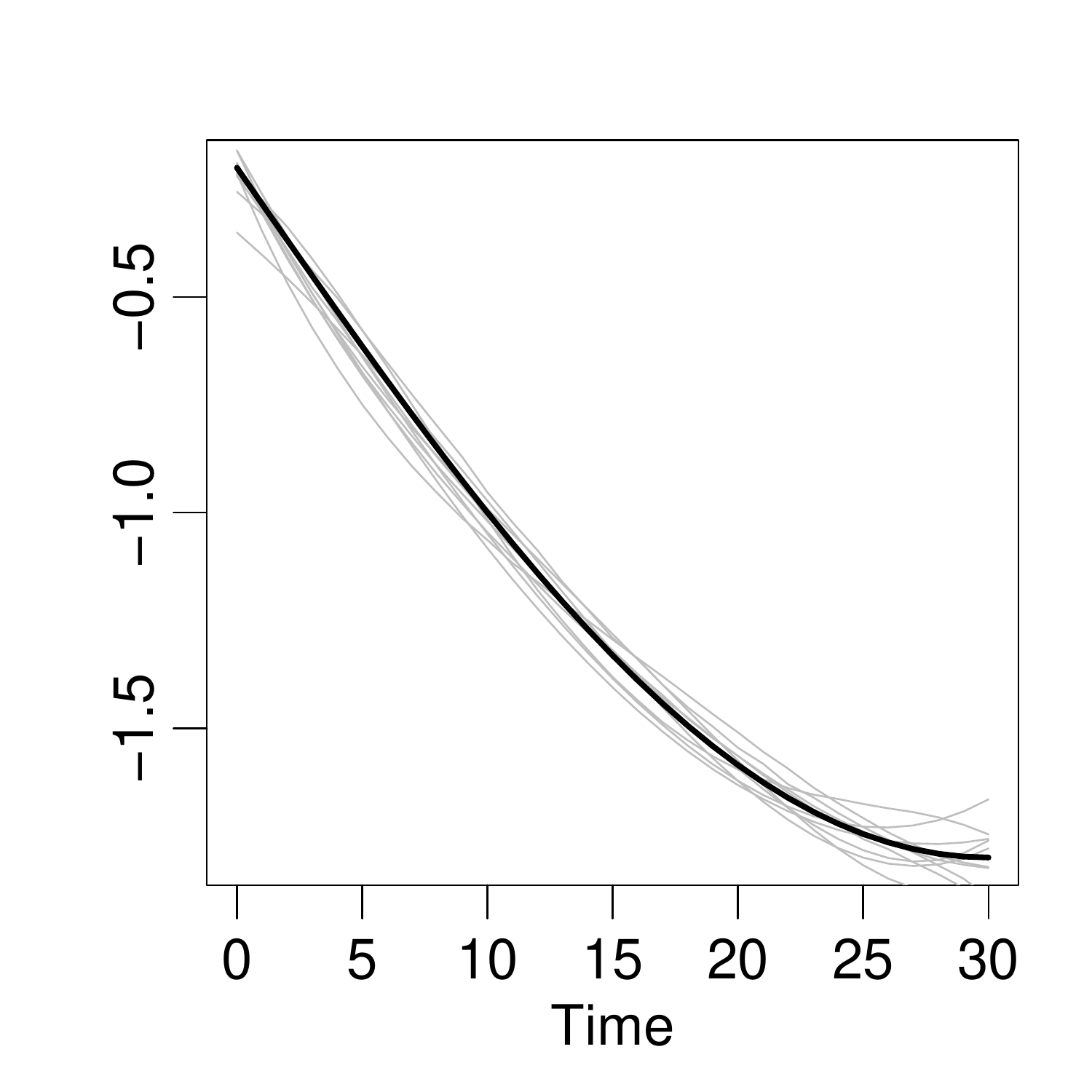} &
		\includegraphics[width=0.33\linewidth]{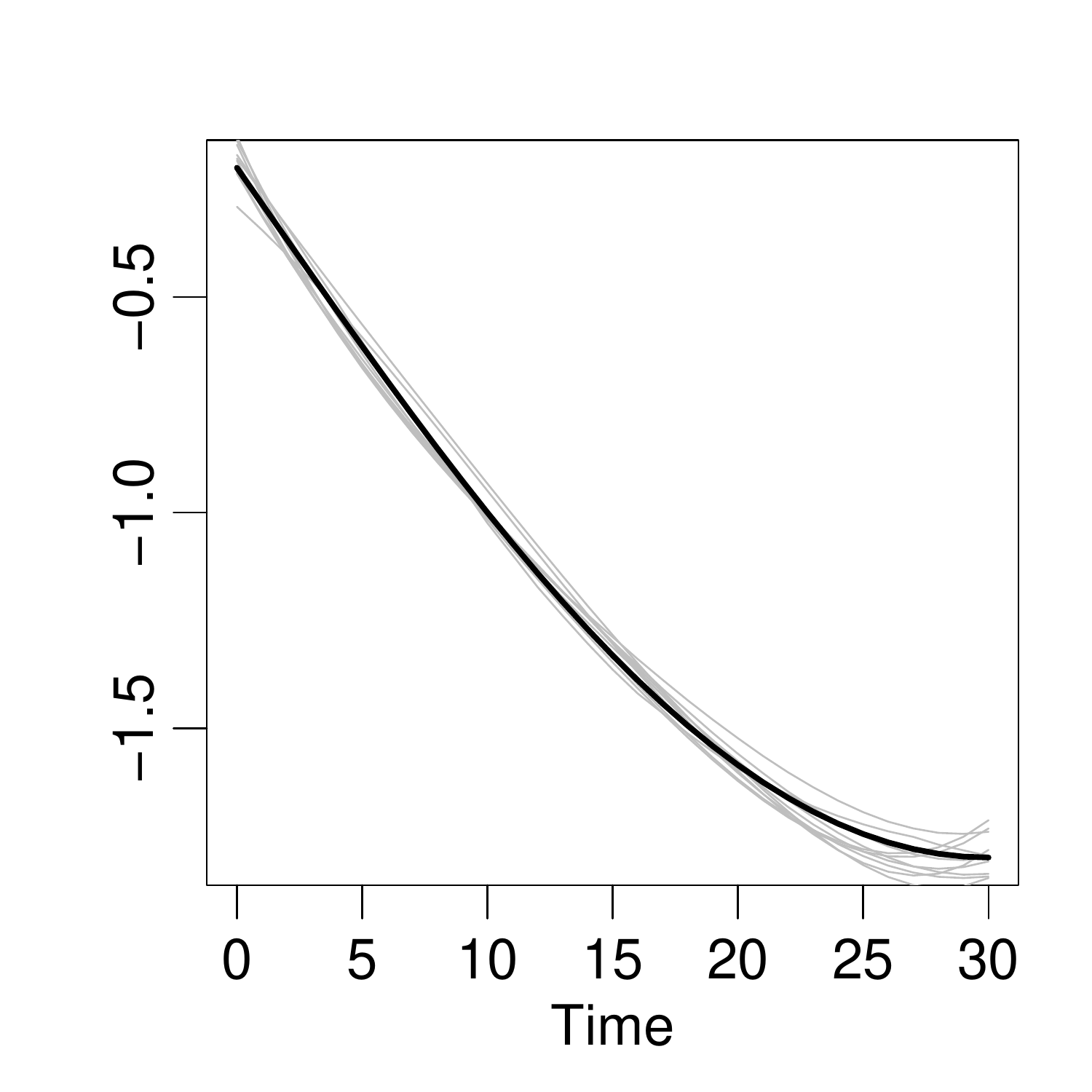} &
		\includegraphics[width=0.33\linewidth]{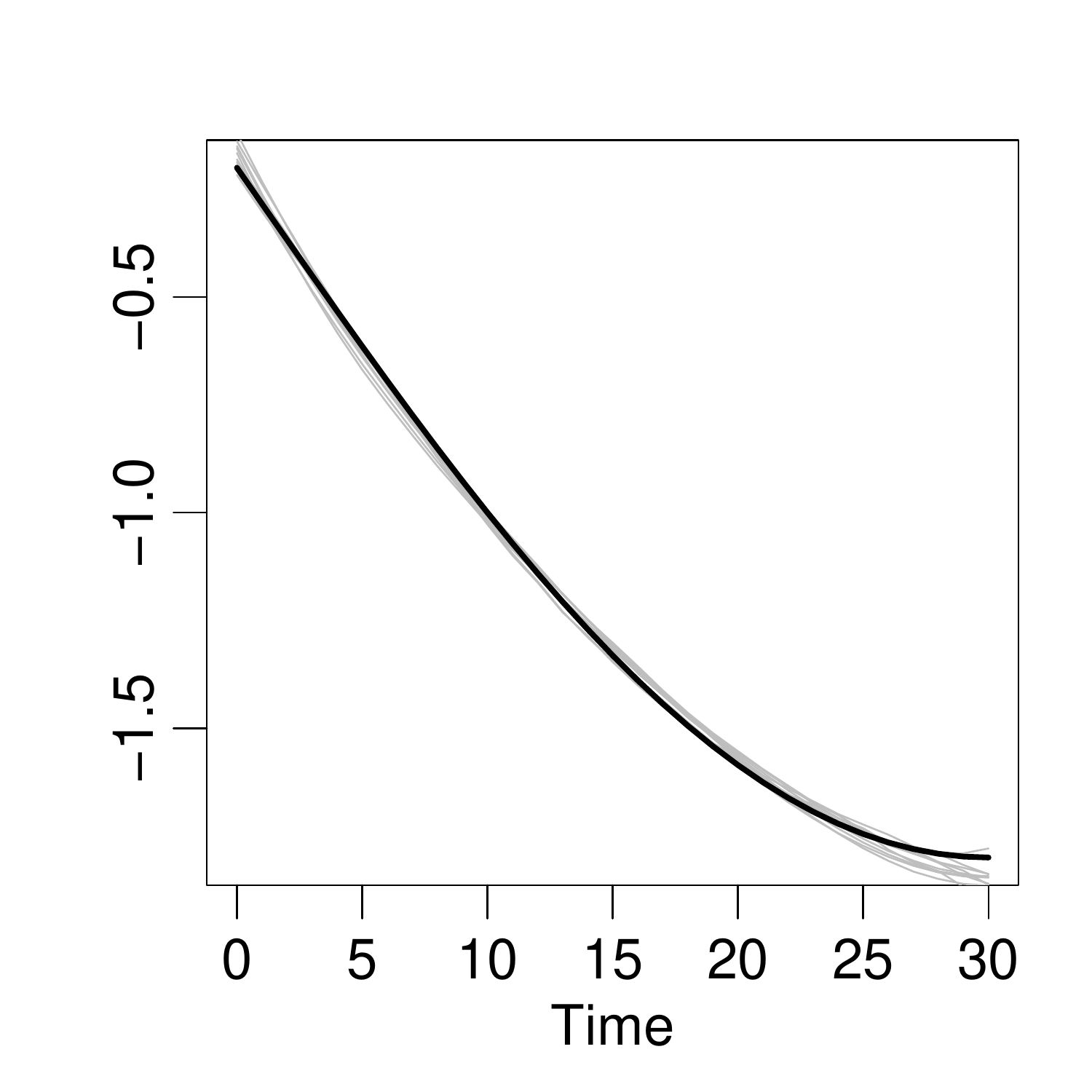} \\
		\begin{sideways} \hspace{55pt} $\beta_2(t)$ \end{sideways}   &
		\includegraphics[width=0.33\linewidth]{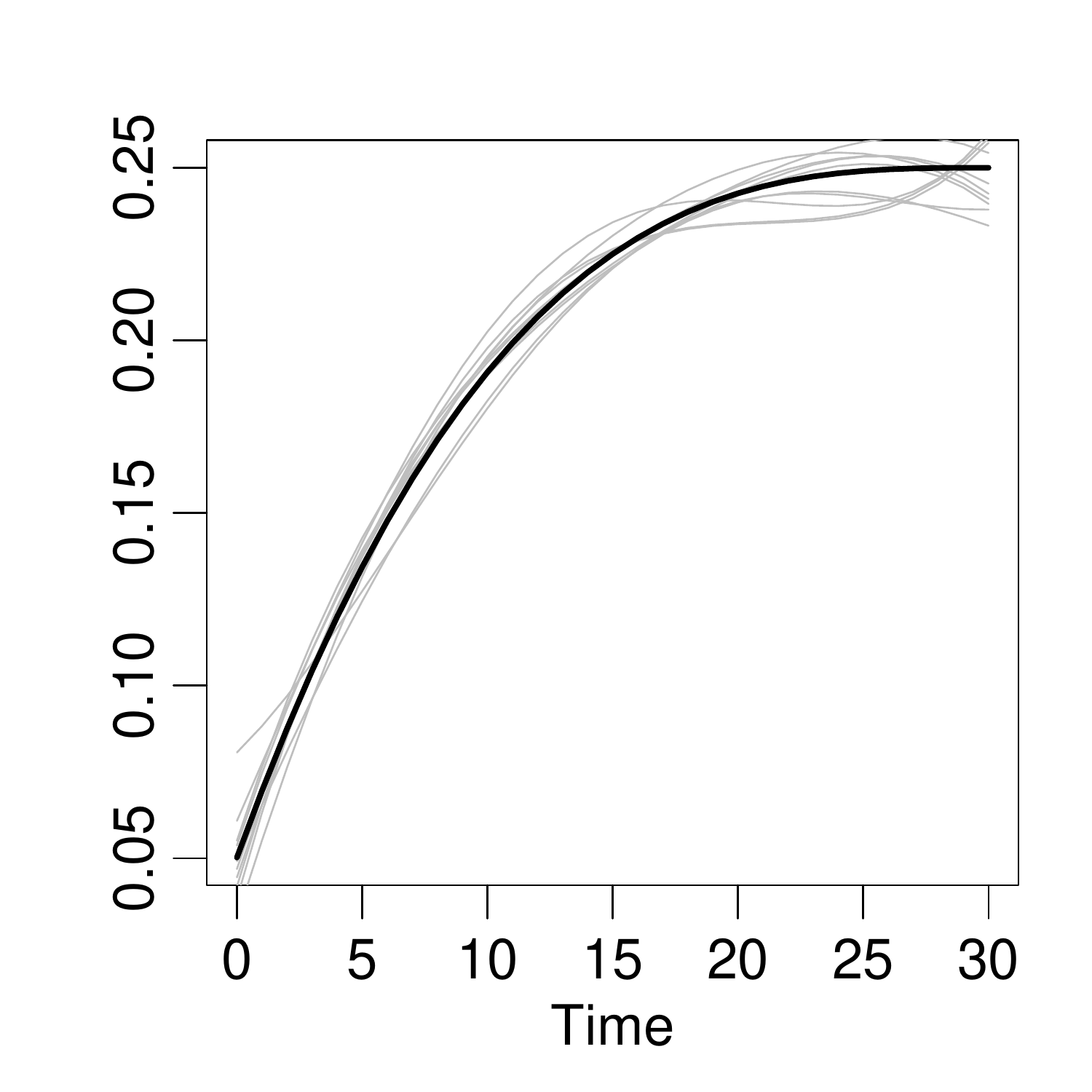} &
		\includegraphics[width=0.33\linewidth]{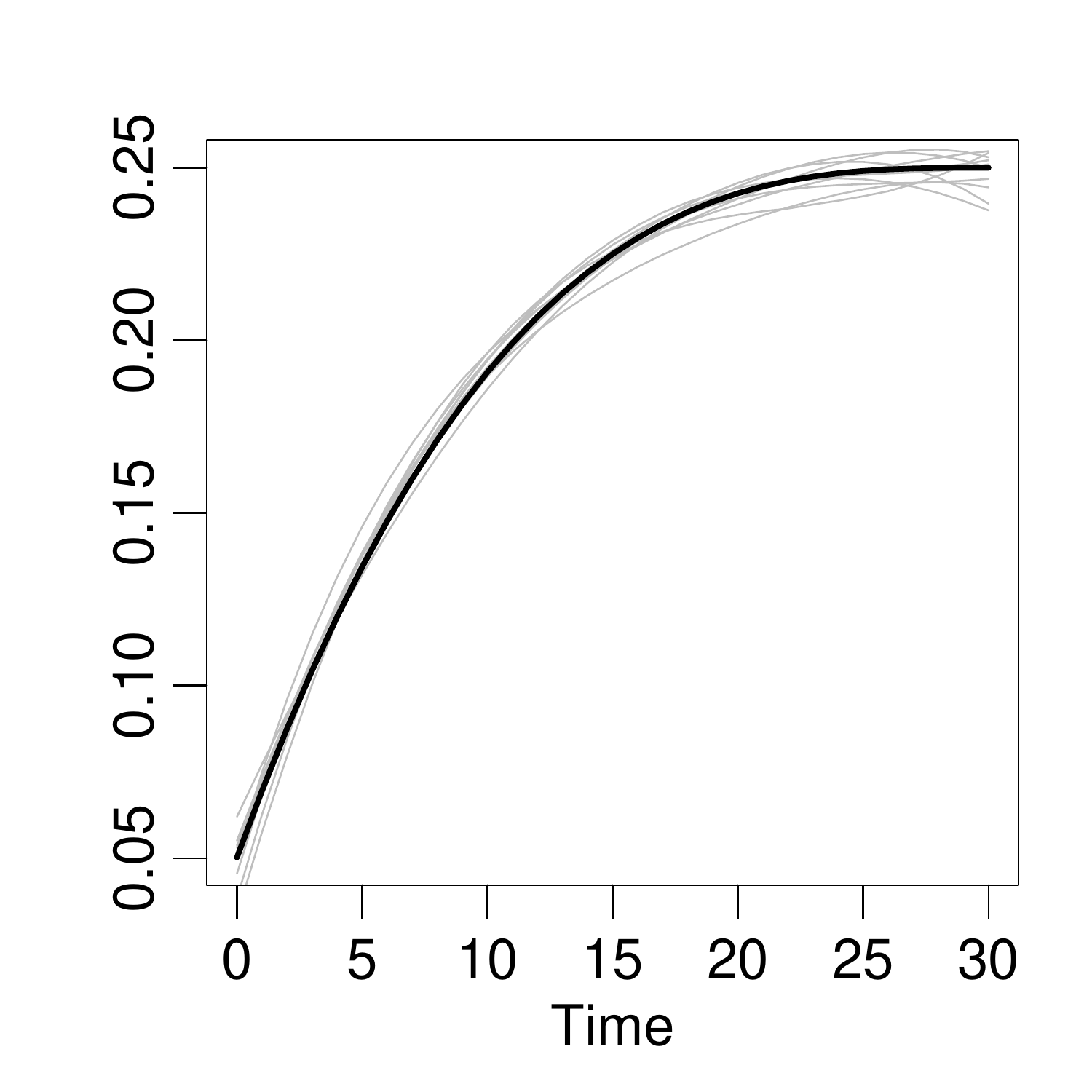} &
		\includegraphics[width=0.33\linewidth]{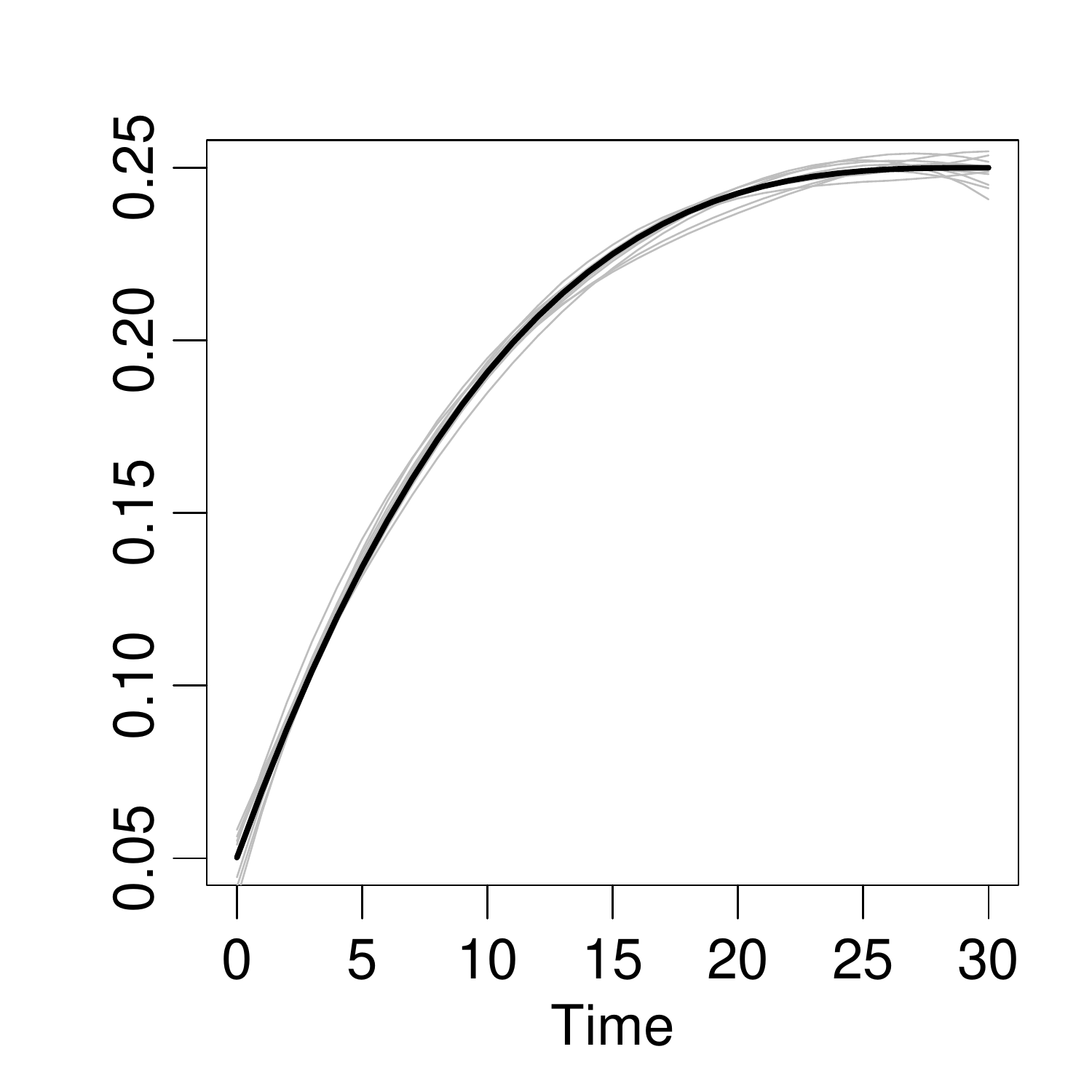} \\
	\end{tabular}
	\caption{\textsf{MADE} distribution and dynamic parameter Bayesian estimates of $\beta_0(t)$, $\beta_1(t)$, and $\beta_2(t)$ (bold lines are the true coefficient functions and gray lines correspond to ten randomly selected estimates), corresponding to 250 synthetic datasets generated according to \textsf{TVCM} \eqref{eq_tvcm_sim3}.  Sample sizes are displayed in columns ($n=25$, $n = 50$, and $n=100$). The model is fitted each time using both radial kernel (\textsf{K}) and regression spline (\textsf{S}) functions, with Frequentist (blue), Bayesian (black), and variational (green) methods.}
	\label{fig_sim3}
\end{figure}

\newpage

In this case, the performance of an estimate is measured by means of the average mean square error (\textsf{AMSE}), which is defined as
\begin{equation*}
\textsf{AMSE}(\beta_r) = \sum_{i=1}^n\sum_{j=1}^{n_i} \frac{1}{n\,n_i}\left(\beta_r(t_{i,j})-\widehat{\beta}_r(t_{i,j})\right)^2\,,\qquad r=0,\ldots,d.
\end{equation*}
Figures \ref{fig_sim1} and \ref{fig_sim2} show the \textsf{AMSE} distribution of $\beta_0(t) = 2e^t$ and $\beta_0(t) =  1 + \cos(2 \pi t) + \sin(2 \pi t)$, respectively, corresponding to 250 synthetic datasets generated according to \textsf{TVCM} \eqref{eq_tvcm_sim1}, in each of nine scenarios delimited by correlation level (weak, medium, and high within-subject correlation) and sample size ($n=25$, $n = 50$, and $n=100$). In general, the \textsf{AMSE} distribution is quite consistent across inference paradigms, which is particularly evident in the first case. Such a behavior was somewhat predictable because the number of measurements, even for the smallest datasets, is big enough (about 375 observations) to allow the likelihood to overcome the prior distribution. Even though \textsf{AMSE}s are also very similar across basis functions, error rates are slightly smaller in the second case when Bayesian methods along with radial functions are employed.

Furthermore, Frequentist inferences are equivalent regardless of the smoothing approach. Also, we observe that the variational approximation to the posterior distribution under the Bayesian paradigm is very precise. This is the case because the mean field assumption is breaking negligible correlations in the posterior distribution. Moreover, as expected, estimates are more consistent as the sample size increases since the variability of the \textsf{AMSE} distribution decreases and its center remains stable. On the contrary, such variability increases as the within-subject correlation becomes higher, which strongly suggests that despite our approach not taking into account within-subject correlation directly, estimates are robust enough to produce accurate results.

\subsection{Simulation scenario 2}

In this experiment, we generate synthetic datasets as follows:
\begin{equation}\label{eq_tvcm_sim3}
y_i(t) = \beta_0(t) + \beta_1(t)x_{1,i} + \beta_1(t)x_{1,i} + \epsilon_i(t)\,,\qquad i = 1,\ldots,n,
\end{equation}
where
\begin{align*}
\beta_0(t) &= 3.5+6.5\sin(t\pi/60)\,,\\
\beta_1(t) &= -0.2-1.6\cos((t-30)\pi/60)\,,\text{ and } \\
\beta_3(t) &= 0.25 -0.0074((30-t)/10)^3 
\end{align*}
are three nonlinear time-varying coefficients, associated with time-invariant independent covariates $x_{1,i}$ and $x_{2,i}$ such that $x_{1,i}\simiid\Ber(0.5)$ and $x_{2,i}\simiid\Nor(0, 4^2)$, and finally, $\epsilon_i(t)$ is Gaussian process with zero mean and covariance function $\gamma(s,t) = 0.0625\,\exp(-|s-t|)$. Also, we assume that subjects are independent of each other and scheduled to be observed at $m=31$ equally spaced time design points $0, 1, \ldots , m$. However, at each given time point, a subject has 50\% probability to be randomly missing. As before, we generated $250$ datasets with three sample sizes, $n=25$, $n=50$, and $n=100$, and once again, we fitted the model \eqref{eq_tvcm} using {\color{black} exactly the same settings as in Section \ref{sec_case_study_1}.}

In order to measure the performance of an estimate fairly, we define the mean absolute deviation of errors (\textsf{MADE}) as 
\begin{equation*}
\textsf{MADE} = \sum_{r=0}^d \sum_{i=1}^n\sum_{j=1}^{n_i} \frac{1}{n\,n_i} \frac{|\beta_r(t_{i,j})-\widehat{\beta}_r(t_{i,j})|}{\text{range}(\beta_r)}\,,\qquad r=0,\ldots,d.
\end{equation*}
Figure \ref{fig_sim3} shows the \textsf{MADE} distribution along with the dynamic parameter Bayesian estimates of the coefficients, corresponding to 250 synthetic datasets generated according to \textsf{TVCM} \eqref{eq_tvcm_sim1}, in each of three scenarios delimited by sample size ($n=25$, $n = 50$, and $n=100$). Again, it is quite obvious the effect of the sample size on the error rates. This behavior is evident from both the decreasing variability of the \textsf{MADE} distribution and the consistent estimates of the coefficients around their true value. Clearly, these simulation results demonstrate that both estimation approaches independently of the inference paradigm, provide reasonably good estimators, at least for interior time points.

{\color{black}
\subsection{Execution time}

\begin{table}[!b]
	\centering
		\begin{tabular}{lcccccc}  
			\hline
			\multicolumn{1}{c}{} & \multicolumn{2}{c}{$n=25$} & \multicolumn{2}{c}{$n=50$} & \multicolumn{2}{c}{$n=100$} \\ \cmidrule{2-7}
			Setting & \textsf{MC} & \textsf{V} & \textsf{MC} & \textsf{V} & \textsf{MC} & \textsf{V} \\ 
			\hline
			Scenario 1a & 128.60 &  7.93 & 160.17 &  6.07 & 347.17 &  7.13  \\
			Scenario 1b & 151.30 & 10.33 & 204.10 &  7.10 & 458.77 &  9.87  \\
			Scenario 2  & 280.43 & 12.03 & 381.67 & 12.37 & 581.30 & 12.77  \\
			\hline
		\end{tabular}
	\caption{Mean running times (in milliseconds) using a single core of an AMD A12-9730P processor, when generating 2,000 samples of the posterior distribution for the model based on radial functions (our proposal) using both MCMC (\textsf{MC}) and variational (\textsf{V}) methods, for each synthetic dataset under all simulation settings considered in Section \ref{sec_simulation}.}\label{tab_times}
\end{table}

Following a suggestion given by one of the referees, here, we provide a comparison in terms of execution time between our two competing approaches to carry out Bayesian inference, namely, MCMC methods and variational methods, i.e., simulation-based methods and optimization-based methods.

In this spirit, Table \ref{tab_times} contains mean running times (in milliseconds) using a single core of an AMD A12-9730P processor, when generating 2,000 samples of the posterior distribution for the model based on radial kernel functions (our proposal) using both MCMC and variational methods, for each synthetic dataset under all simulation settings considered in Section \ref{sec_simulation}.  We see that the variational approach clearly greatly outperforms its simulation-based counterpart in terms of execution time. Such an effect is particularly clearer for bigger samples sizes, where variational methods can even be 45 faster than MCMC methods. Lastly, note that	 MCMC execution times increase notoriously as the sample size grow, whereas variational execution times remain roughly constant.
}

\section{Illustrations}\label{sec_illustrations}

\subsection{Case study 1}\label{sec_case_study_1}

Our first illustration is based on an AIDS clinical trial developed by the AIDS Clinical Trials Group\footnote{Visit \url{https://actgnetwork.org/} for more information about the group.} (ACTG). In this group, \cite{fischl-03} evaluated two different 4-drug regimens containing \textit{indinavir} with either \textit{efavirenz} or \textit{nelfinavir} for the treatment of 517 patients with advanced HIV disease (i.e., patients with high HIV-1 RNA levels and low CD4 cell counts). This study was a randomized, open-label study and initially planned to last 72 weeks but later increased to 120 weeks beyond the enrollment of the last subject. The randomization was carried out by using a permuted block design and stratified according to CD4 cell count and HIV-1 RNA level at screening, as well as previous antiretroviral experience. In addition, clinical assessments, HIV-1 RNA measurements, CD4 cell counts, and routine laboratory tests were performed before study entry, at the time of study entry, at weeks 4 and 8, and every 8 weeks thereafter. More details about design, subjects, treatments and outcome measurements of this study are given in \cite{fischl-03}.

\begin{figure}[!h]
	\centering
	\includegraphics[scale=0.4]{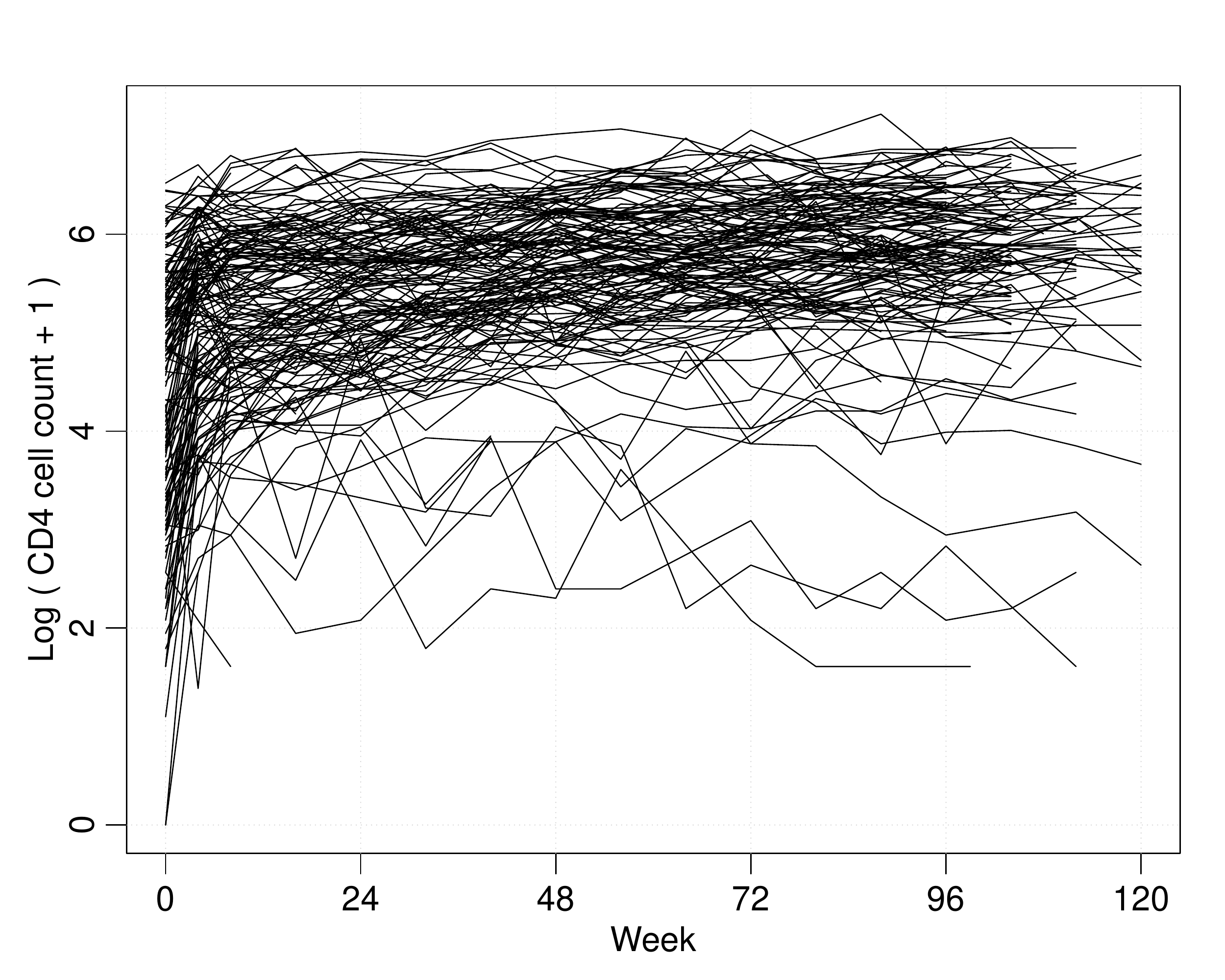}
	\caption{ACTG 388 data: CD4 cell counts (logarithmic scale).}\label{fig_spaghetti_plot_ACTG_388}
\end{figure}

Here, we model the CD4 cell count, which is an essential marker for assessing immunologic response of an antiviral regimen, in one of the two treatment arms. This group includes 166 patients treated with highly active antiretroviral therapy for 120 weeks, during which CD4 cell counts were monitored along with other important markers. Patients might not exactly follow the designed schedule, and missing clinical visits for CD4 cell measurements frequently occurred, which makes this dataset\footnote{\url{http://www.urmc.rochester.edu/biostat/people/faculty/wusite/datasets/data/ACTG388Data1Arm.cfm}.} (named ACTG 388) a typical longitudinal dataset. The main goal in this study is to model the mean CD4 cell count trajectories over the treatment period for the entire treatment arm.

In this specific group of patients, the number of CD4 cell count measurements per patient varies from 1 to 18 observations, and the CD4 cell count ranges from 0 to 1,364. Figure \ref{fig_spaghetti_plot_ACTG_388} shows CD4 cell counts (in logarithmic scale) for each one of the $n=166$ patients during the 120 weeks of treatment. Even though individual cell counts are quite noisy and there is evidence of some atypical trajectories associated with low counts, this plot suggests that cell counts tend to stabilize around the middle of the treatment. Thus, it is not possible to ensure that the antiviral treatment was quite effective since there are no apparent reasons to believe that CD4 cell counts profiles are either increasing continuously or at least remaining stable.

In order to estimate the mean trajectory of CD4 cell counts over the treatment period, we fit the TVCM
$$y_{i,j} =\beta_0(t_{i,j}) + \epsilon_{i,j}\,, \qquad j=1,\ldots,n_i\,,\qquad i=1,\ldots,n\,,$$ 
under a Bayesian {\color{black} approach with the prior distribution given in Section \ref{sec_bayesian_inference},} employing both Gaussian kernel and regression spline functions with $g=2$, where $y_{i,j}$ is the CD4 cell count (in logarithmic scale) of the $j$-th measurement of the $i$-th patient, and $\beta_0(t)$ is a unknown time-varying parameter describing the mean dynamic trend of CD4 cell counts over time. Again, {\color{black} the number and location of knots are chosen according to the \textsf{PCV} criteria and the equally spaced method described in Section \ref{sec_selection_number_location_knots}, respectively.}. According to this criteria, the optimal number of knots are $k_0^{\textsf{K}} = 4$ and $k_0^{\textsf{S}} = 8$, respectively. {\color{black} Once the number and location of knots are fixed, Bayesian estimates are based on $2,000$ samples from the posterior distribution after a burn-in period of 500 iterations.}  Convergence was monitored by tracking the variability of the joint distribution of data and parameters using the multi-chain procedure discussed in \cite{GeRu92}.

\begin{figure}[!h]
	\centering
	\includegraphics[scale=0.34]{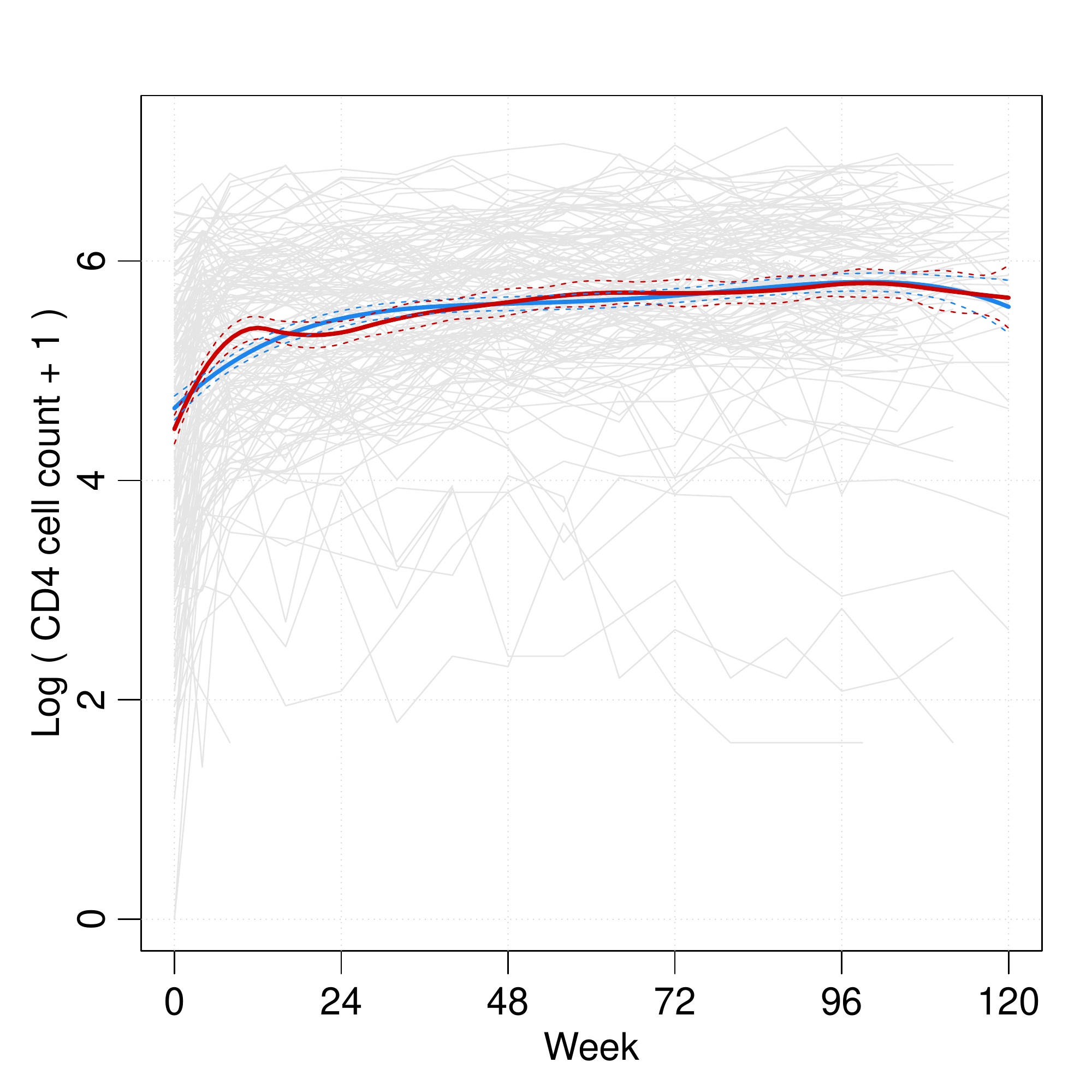}
	\includegraphics[scale=0.34]{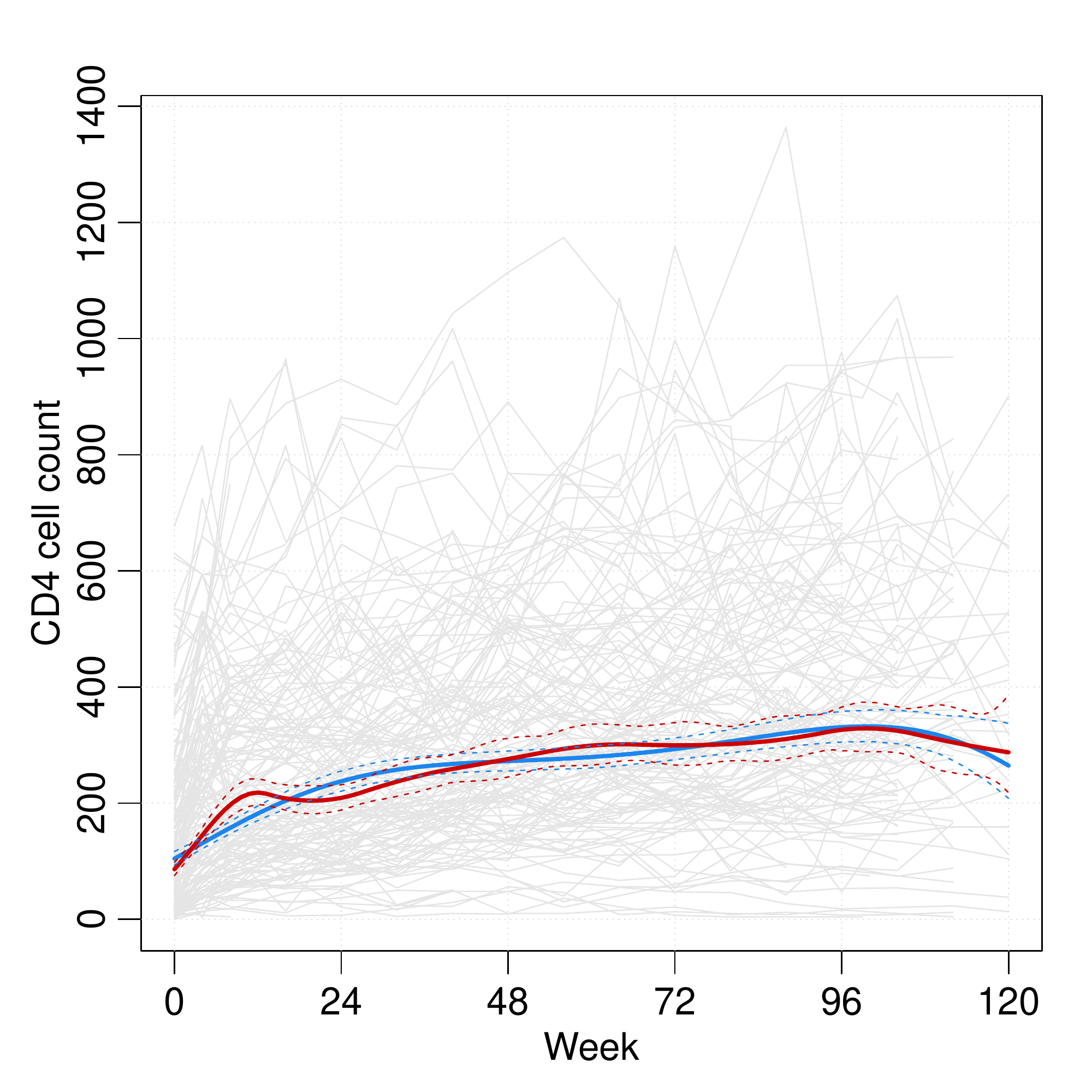}
	\caption{$\beta_0(t)$ estimates (solid lines) along with 95\% credible intervals (dotted lines) obtained through radial kernel functions (blue) and regression spline functions (black). Individual trajectories are displayed in gray.}\label{fig_fitted_curve_ACTG_388}
\end{figure}

Estimates of $\beta_0(t)$ (in logarithmic and natural scale) along with their corresponding 95\% credible intervals are shown in Figure \ref{fig_fitted_curve_ACTG_388}. Both estimates are very similar, except for the small jump at the beginning of the treatment exhibited by the regression spline-based estimate. Such trajectories, which are very precise since the credible intervals are quite narrow, reveal that the mean CD4 cell counts increase quite sharply during the first 40 weeks of treatment, and continue to increase at a slower rate until about week 100, and then dropped towards the end of the study. This makes evident that under this antiviral regimen, the overall CD4 counts increased dramatically during the first 40 weeks, but the effect of the drug therapy fades over time and completely disappeared after about week 100, when cell counts begin to drop. 
Almost identical results were obtained by \cite{wu-zhang-06}; the only difference is that they concluded that the inflection point after which the CD4 cell count started to drop was on week 110. A residual analysis (not shown here) indicates that the model
fits the data adequately because there are no signs of particular shapes, patterns or significant deviations.

\subsection{Case study 2} \label{sec_case_study_2}

\begin{figure}[!t]
	\centering
	\includegraphics[scale=0.4]{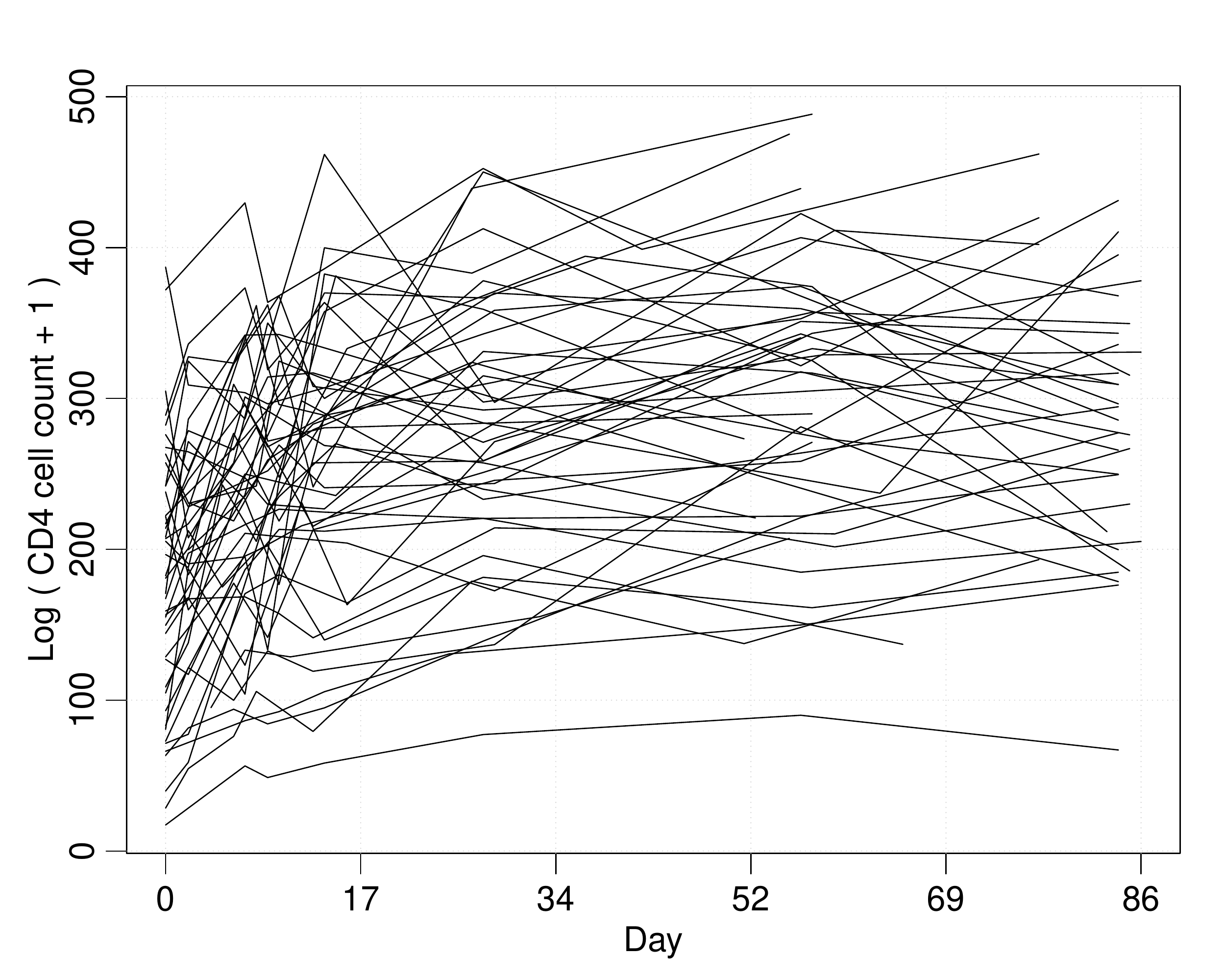}
	\caption{ACTG 315 data: CD4 cell counts (logarithmic scale).}\label{fig_spaghetti_plot_ACTG_315}
\end{figure}

We consider another AIDS clinical study carried out by the ACTG. In this case, \cite{lederman-98} evaluated a highly active antiretroviral therapy containing \textit{zidovudine}, \textit{lamivudine}, and \textit{ritonavir}, for the treatment of patients with moderately advanced HIV-1 infection. This study was designed to ascertain if administration of highly active antiretroviral therapy to patients with moderately advanced HIV-1 infection was associated with evidence of immunologic restoration. More details about design, subjects, treatments and outcome measurements of this study are given in \cite{lederman-98}.

The viral load (plasma HIV RNA level) and immunologic response (CD4 cell counts) are negatively correlated and their relationship is approximately linear during antiviral treatments. However, their relationship may not be a constant during the whole period of treatment \cite{liang-2003-relationship}. Thus, the main goal in this study is to model the dynamic relationship between the viral load and the immunologic response over the treatment period, which plays an essential role in evaluating the antiviral therapy. Fifty-three patients were enrolled in the trial, out of which $n=46$ received the treatment for at least 9 of the first 12 weeks and were therefore eligible for analysis. Intolerance of the treatment regimen was responsible for almost all treatment discontinuations. Patients might not exactly follow the designed schedule, and missing clinical visits occurred frequently, which makes this dataset\footnote{\url{http://www.urmc.rochester.edu/biostat/people/faculty/wusite/datasets/data/ACTG315LongitudinalDataViralLoadData.cfm}.} (named ACTG 315) quite unbalanced. Additional analyses of this and other trajectories, as well as more scientific findings of the study, can be found in \cite{lederman-98}, \cite{connick-00}, \cite{liang-2003-relationship}, and \cite{wu-2004-backfitting}.

\begin{figure}[!t]
	\centering
	\includegraphics[scale=0.35]{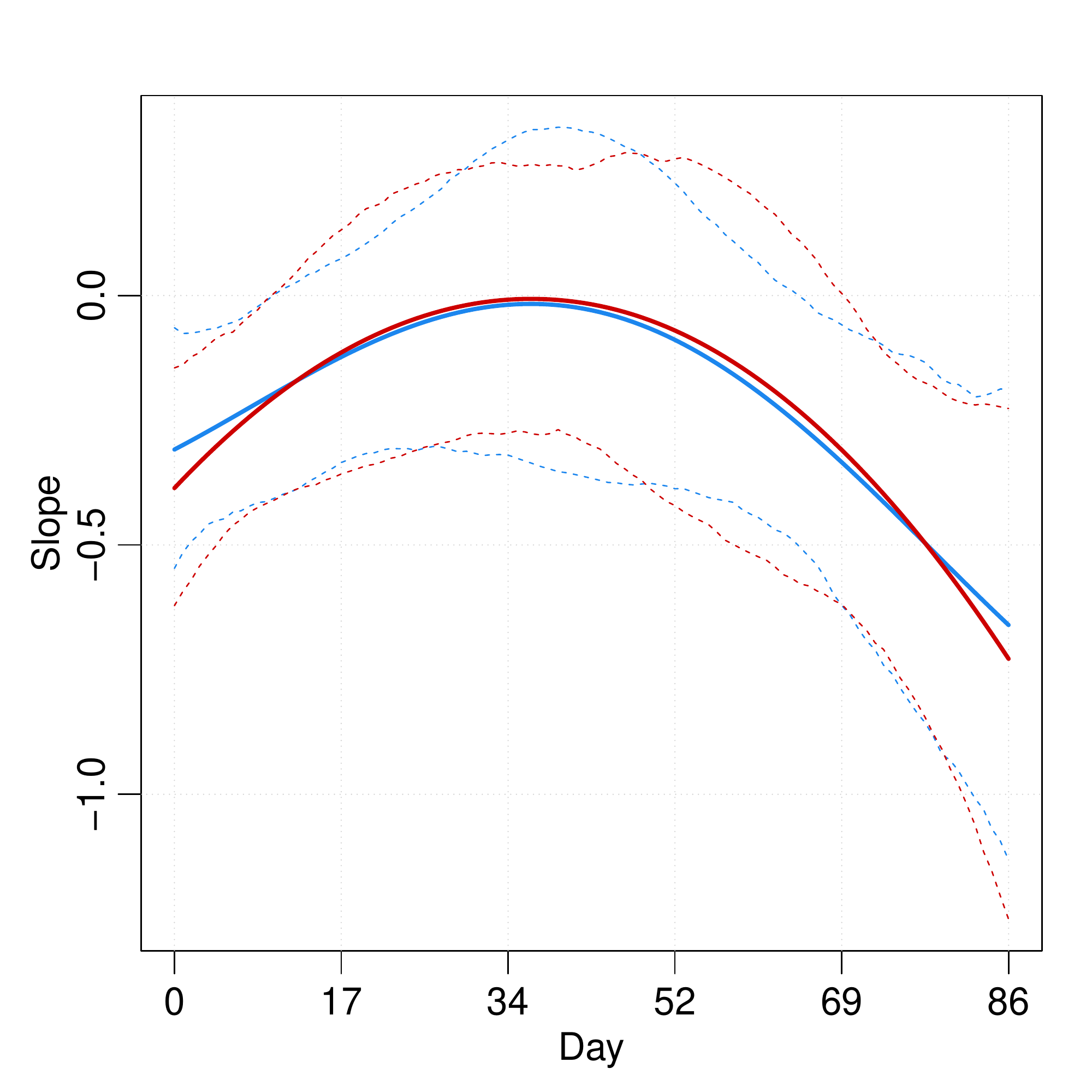}
	\caption{$\beta_1(t)$ estimates (solid lines) along with 95\% credible intervals (dotted lines) obtained through radial kernel functions (blue) and regression spline functions (black).}\label{fig_fitted_curve_ACTG_315}
\end{figure}

After starting treatment, the plasma HIV RNA level and the CD4 cell count were measured simultaneously (both of them were reported in logarithmic scale) at days 0, 2, 7, 10, 14, 28, 56, and 86. Figure \ref{fig_spaghetti_plot_ACTG_315} shows the corresponding cell counts. The number of repeated measurements per subject varies from 4 to 8, and the total number of observations is 328. Simple linear regressions of cell counts against plasma HIV RNA level at each visit (not shown here) evidence that the slope associated with the viral load changes over time because some days the slope is significantly different from zero. This simple observation motivates fitting a TVCM in order to characterize and quantify such relationship.

Once again, under a Bayesian setting, employing both Gaussian kernel and regression spline functions with $g=2$, we fit the TVCM given by
$$
y_{i,j} =\beta_0(t_{i,j}) + \beta_1(t_{i,j})x_{1,i}(t_{i,j}) + \epsilon_{i,j}, \,, \qquad j=1,\ldots,n_i\,,\qquad i=1,\ldots,n\,,
$$
where $y_{i,j}$ and $x_{1,i}(t_{ij})$ are the viral load (in logarithmic scale) and the CD4 cell count (also in logarithmic scale) associated with the $j$-th measurement of the $i$-th patient, respectively. The time-varying coefficient $\beta_1(t)$ characterizes the dynamic relationship between the viral load and the immunologic response over the treatment period. Interestingly, according to \textsf{PCV} criteria, the optimal number of knots in both cases are $k_0^{\textsf{K}} = k_0^{\textsf{S}} = k_1^{\textsf{K}} = k_1^{\textsf{S}} = 1$. {\color{black} Following exactly the same setting as in the case study \#1,} the results we report are based on 2,000 samples obtained after a burn-in period of 500 iterations.

Figure \ref{fig_fitted_curve_ACTG_315} shows estimates of the dynamic slope. Both estimated trajectories are quite smooth. Here, we see a significant negative correlation between viral load and immunologic response at the beginning of the treatment. Then, the relationship consistently attenuates until reaching zero about the fourth week. At that point, the negative correlation gradually strengthens again, and continuously to do so towards the end of the treatment period considered in this analysis. Almost identical results were obtained by \cite{wu-2004-backfitting}.

\subsection{Goodness-of-fit and predictive performance}

The modeling literature has largely focused on both Akaike Information Criteria (\textsf{AIC}) and Bayesian Information Criteria (\textsf{BIC}) as a tool for model selection (e.g., see \citealp{wu-zhang-06}). However, under a Bayesian setting, the \textsf{BIC} is inappropriate for hierarchical models since it underestimates the complexity of the model.  An alternative to \textsf{BIC} that addresses this issue is the Deviance Information Criterion (\textsf{DIC}),
$\textsf{DIC} = - 2\log p ( \yv \mid \widehat{\UPS} ) + 2p_{\textsf{DIC}}$,
where $\widehat{\UPS}$ is the posterior mean of model parameters and $p_{\textsf{DIC}} = 2 \log p ( \yv \mid \widehat{\UPS} ) - 2\, \expec{\log p \left(\yv \mid \UPS \right) }$ is the model complexity \citep[see][for a discussion]{gelman-2013-bayesian}.
Table \ref{tab_datasets_amse} presents the \textsf{DIC} for the TVCMs fitted above. It is clear the the TVCM based on radial kernel functions is preferred, specially for the ACTG 388 dataset.

\begin{table}[!h]
	\centering
	\begin{tabular}{lcccc}  
		\hline
		\multicolumn{1}{c}{} & \multicolumn{2}{c}{Kernel} & \multicolumn{2}{c}{Spline} \\ \cmidrule{2-5}
		Dataset  & \textsf{DIC} & \textsf{AMSE} & \textsf{DIC} & \textsf{AMSE} \\ 
		\hline
		ACTG 388 & 64,257.1  & 0.761  & 65,306.7  & 0.759 \\
		ACTG 315 & 2,960.9   & 0.758  & 2,955.8   & 0.811 \\
		\hline
	\end{tabular}
	\caption{\textsf{DIC} and \textsf{AMSE} to assess goodness-of-fit and out-sample predictive performance, respectively, of both radial kernel and regression spline functions.}\label{tab_datasets_amse}
\end{table}

On the other hand, in order to compare the ability of each alternative to predict missing observations, we evaluate their out-of-sample predictive performance by means of a cross-validation (CV) experiment. Thus, for each combination of dataset and model, we performed an $L$-fold CV in which $L$ randomly selected subsets of roughly equal size in the dataset are treated as missing and then predicted using the rest of the data. We summarize our findings in Table \ref{tab_datasets_amse}, where we report the average mean square error (\textsf{AMSE}) corresponding to the prediction of missing measurements in the datasets.  In this context, the \textsf{AMSE} is a measure of how well a given model is capable of predicting missing observations. We can see from this table that both alternatives have comparable predictive capabilities.

\section{Discussion}\label{sec_discussion}

In this paper, we review two simple but powerful alternatives based on linear expansions to estimate time-varying coefficients: a new approach using radial kernel functions and a more standard method using regression spline functions. We framed the estimation procedure under both Frequentist and Bayesian inference paradigms using bootstrap techniques, Gibbs sampling and a variational Bayes method. From an empirical perspective, we provide two simulations studies. These experiments strongly suggest that either combination of basis representation and inference approach are comparable and mostly equivalent. From a practical perspective, the first case study shows that the overall CD4 counts increased dramatically during the first 40 weeks under a specific anti-viral treatment, but the effect of the drug therapy faded over time and completely disappeared after about week 100. On the other hand, the second case study evidences a strong negative correlation between the viral load and the immunologic response at the beginning of an anti-viral treatment, and then shows evidence of a weak correction about the fifth week, when gradually strengthened again and reached the largest value at the end of the treatment period.

{\color{black} As part of the revision process, one of the referees suggested that, in order to avoid inconsistencies in terms of exposition, the model needed to be presented in a general fashion as in a mixed-effects time-varying coefficient model (see Section \ref{sec_simulation_1}). Even though we agree on the convenience of working with a more general model, we consider that our approach should be treated in terms of a ``standard'' \textsf{TVCM} as in Equation (1) since it makes exposition straightforward, given that our main contribution rely on an estimation protocol based on radial functions, along with inference strategies according to the Bayesian paradigm.  Nonetheless, we exhort the reader to pursue such an extension employing the ideas discussed in this manuscript.}

The estimation protocol presented here is susceptible of many extensions. For instance, to avoid the curse of dimensionality, the model can be extended to account for longitudinal inhomogeneity of varying coefficients via Bayesian basis selection or adaptive knot selection which is an integral part of the data generating mechanism. Another interesting extension involves the incorporation of specific working correlation matrices in the probabilistic structure of the random error using more convoluted covariance functions. Extensions to more complex situations, including multivariate or spatial data are also possible, and are the subject of future work.

\bibliographystyle{apalike}
\bibliography{references}

\appendix

\section{Algortihms for Inference }

\subsection{Bootstrap algorithm}\label{app_bootstrap}

Since all subjects are assumed to be independent, a natural sampling scheme consists in resampling the entire repeated measurements of each subject with replacement from the original dataset. The bootstrap samples $\Upsv^{(1)},\ldots,\Upsv^{(B)}$, where $\Upsv^{(b)} = \left( \alv^{(b)},\sig^{(b)} \right)$, $b = 1,\ldots,B$, can be generated using the following bootstrap algorithm:
\begin{enumerate}[1.]
	\item Randomly select $n$ bootstrap subjects with replacement from the original dataset, and put together the bootstrap dataset as $\mathcal{D}^{(b)} = \left\{(y^*_{i,j}, \xv^*_{i,j}, t^*_{i,j}, ) : j=1,\ldots,n_i, i=1,\ldots,n\right\}\,.$
	Note that the entire set of repeated measurements for some subjects may appear multiple times in the bootstrap dataset.
	\item Compute $N^{(b)}$, $\yv^{(b)}$, $\Z^{(b)}$, and $\W^{(b)}$ based on $\mathcal{D}^{(b)}$.
	\item Compute $\alv^{(b)}$   as in \eqref{eq_WLS_alpha_estimator} based on $\yv^{(b)}$, $\Z^{(b)}$, $\W^{(b)}$.
	\item Compute $\sigma^{(b)}$ \,as in \eqref{eq_sig2_estimator}    based on $\yv^{(b)}$, $\Z^{(b)}$, $\W^{(b)}$, $\alv^{(b)}$.
	\item Repeat the previous steps $B$	times.
\end{enumerate}

\subsection{MCMC algorithm}\label{app_mcmc}

The posterior distribution of the parameters can be explored using MCMC algorithms in which the posterior distribution is approximated using dependent but approximately identically distributed samples $\Upsv^{(1)},\ldots,\Upsv^{(B)}$, where $\Upsv^{(b)} = \left( \alv^{(b)},\sig^{(b)} \right)$, $b = 1,\ldots,B$. 
In this case, the joint posterior distribution is given by:
\begin{align}
p(\alv,\sig^2\mid\Z,\yv) 
&\propto (\sig^2)^{-N/2}\,\ex{-\tfrac{1}{2\sig^2}(\yv-\Z\alv)^\trans(\yv-\Z\alv)} \nonumber \\
&\hspace{1cm} \times (\sig^2)^{-p/2}\,\ex{-\tfrac{1}{2N\sig^2}\alv^\trans\alv}
\times (\sig^2)^{-(a_\sig + 1)}\,\ex{-\tfrac{b_\sig}{\sig^2}}\,,  \label{eq_posterior}
\end{align}
where $p=\sum_{r=0}^d p_r$ is the expansion dimension. The MCMC algorithm iterates over the full conditional distributions of the model parameters $\alv$ and $\sig^2$ by generating a new state $\Upsv^{(b+1)}$ from a current state $\Upsv^{(b)}$ as follows:
\begin{enumerate}
	\item Choose an initial value for $\alv$, say $\alv^{(0)}$.
	\item Update $(\sig^2)^{(b+1)}$ and $\alv^{(b+1)}$ until convergence:
	\begin{enumerate}[i.]
		\item Sample $(\sig^2)^{(b+1)}$ from
		$$
		\IGamd\left(a_\sig + \tfrac{N}{2} + \tfrac{p}{2} , b_\sig + \tfrac12\|\yv-\Z\alv^{(b)}\|^2 + \tfrac{1}{2N}\|\alv^{(b)}\|^2\right)\,.
		$$
		\item Sample $\alv^{(b+1)}$ from
		$$
		\Nor\left( \left(\Z^\trans\Z + \tfrac{1}{N}\I_p\right)^{-1} \Z^\trans\yv , (\sig^2)^{(b+1)}\left(\Z^\trans\Z + \tfrac{1}{N}\I_p\right)^{-1} \right)\,.
		$$
	\end{enumerate}
\end{enumerate}
Posterior summaries along with point and interval estimates can be approximated based on the Monte Carlo samples. As before, for a given $t$, the $100(1-\alpha)\%$ percentile-based credible interval for $\be_r(t)$, $r=0,1,\ldots,d$, can be computed as in \eqref{eq_IC_betas}.

\subsection{Variational Bayes algorithm}\label{app_variational}

The Markov chain defined in the previous section is guaranteed to converge eventually to the posterior distribution $p(\alv,\sig^2\mid\Z,\yv)$ given in \eqref{eq_posterior}. Here, we consider the problem of finding a function $q(\cdot)$ in a family of functions closest to the posterior distribution $p(\cdot)$, according to a given dissimilarity measure. This idea is known as variational Bayes \citep[see][for a review]{ormerod-2010-explaining}.

Briefly, the main idea can be summarized as follows. Let $\tev=(\te_1,\ldots,\te_k)$ be a set of parameters and $p(\tev \mid \X)$ its posterior distribution after data $\X$ have been observed. The purpose consists in finding a function $q(\tev)$ that minimizes the  Kullback-Leibler divergence respect to $p(\tev \mid \X)$:
\begin{equation*}
\text{\textsf{KL}}(p||q)=\int q(\tev)\,\log\frac{q(\tev)}{p(\tev\mid\X)}\,\text{d}\tev = \log p(\X) - \int q(\tev)\log\frac{p(\tev,\X)}{q(\tev)}\,\text{d}\tev\,.
\end{equation*}
Thus, minimizing $\text{\textsf{KL}}(p||q)$ is equivalent to maximizing
$$
\text{\textsf{E}}_{q(\tev)}\left[\log\frac{p(\tev,\X)}{q(\tev)}\right] = \int q(\tev)\log\frac{p(\tev,\X)}{q(\tev)}\,\text{d}\tev\,,
$$
which is known as evidence lower bound (\textsf{ELBO}). Furthermore, if $q(\tev)$ is assumed to satisfy the mean field assumption
$$
q(\tev) = \prod_{i=1}^k q_i(\theta_i)\,,
$$
where the $q_i(\theta_i)$ are marginal variational densities, the solution $q^*(\tev)$ satisfies
$$
\log q_i^*(\te_i) \propto \text{\textsf{E}}_{q(\tev_{-i})} \left[ \log p(\te_i\mid\X,\tev_{-i}) \right]\,,
$$
with $\tev_{-i} = (\theta_1,\ldots,\theta_{i-1},\theta_{i+1},\ldots,\theta_k)$, which leads to a coordinate optimization algorithm.

In this case, the product density approximation to $p(\alv, \sig^2\mid\Z,\yv)$ is
$
q(\alv, \sig^2) = q(\alv)\,q(\sig^2).
$
The optimal densities take the form
$$
\log q(\alv)   \propto \textsf{E}_{q(\sig^2)}\left[ \log p(\alv  \mid\Z,\yv,\sig^2) \right] \qquad\text{and}\qquad
\log q(\sig^2) \propto \textsf{E}_{q(\alv)}  \left[ \log p(\sig^2\mid\Z,\yv,\alv) \right] \,.
$$
Then,
$$
q(\alv)
\propto
\ex{ -\frac{1}{2} \left[ \alv^\trans \textsf{E}_{q(\sig^2)}\left[\sig^{-2}\right]\left(\Z^\trans\Z + \tfrac{1}{N}\I_p  \right)\alv -2\alv^\trans\textsf{E}_{q(\sig^2)}\left[\sig^{-2}\right]\Z^\trans\yv \right] }
$$
which we can recognize as a member of the normal family
$
\Nor\left( \mv^* , \V^* \right)
$
where
$$
\V^*  = \left(\textsf{E}_{q(\sig^2)}\sig^{-2}\right)^{-1}\left( \Z^\trans\Z + \tfrac{1}{N}\I_p \right)^{-1} \qquad\text{and}\qquad
\mv^* = \V^*\left(\textsf{E}_{q(\sig^2)}\sig^{-2}\right)\Z^\trans\yv\,.
$$
Similar arguments lead to
$$
q(\sig^2) 
\propto (\sig^2)^{-(a_\sig+N/2+p/2 + 1)}\, \ex{ -\frac{1}{\sig^2}\left(b_\sig + \tfrac12\textsf{E}_{q(\alv)} \|\yv-\Z\alv\|^2 + \tfrac{1}{2N}\textsf{E}_{q(\alv)} \|\alv\|^2\right) }\,,
$$
which we can recognize as a member of the inverse-gamma family 
$
\IGamd \left( a^* , b^* \right)
$
where
\begin{align*}
a^* &= a_\sig+N/2+p/2 \qquad\text{and}\qquad \\
b^* &= b_\sig + \tfrac12 \left( \|\yv\|^2 -2\yv^\trans\Z\textsf{E}_{q(\alv)}\alv + \textsf{E}_{q(\alv)}\alv^\trans\left( \Z^\trans\Z + \tfrac{1}{N}\I_p \right)\mathbb{E}_{\alv}\alv + \tr\left[\left( \Z^\trans\Z + \tfrac{1}{N}\I_p \right) \textsf{Var}_{q(\alv)}\alv \right] \right)\,.
\end{align*}
Hence, the algorithm for obtaining the parameters in $q(\alv)$ and $q(\sig^2)$ is the following:
\begin{enumerate}
	\item Initialize $b^* > 0$.
	\item Repeat until the increase in \textsf{ELBO} is negligible:
	\begin{enumerate}[i.]
		\item $\V^*  \,\leftarrow \frac{b^*}{a^*}\left( \Z^\trans\Z + \tfrac{1}{N}\I_p \right)^{-1}$
		\item $\mv^* \leftarrow \frac{a^*}{b^*}\,\V^*\,\Z^\trans\yv$
		\item $b^*   \,\,\,\,\leftarrow  b_\sig + \tfrac12 \left( \|\yv\|^2 -2\yv^\trans\Z\mv^* + \mv^{*\trans}\left( \Z^\trans\Z + \tfrac{1}{N}\I_p \right)\mv^* + \tr\left[\left( \Z^\trans\Z + \tfrac{1}{N}\I_p \right) \V^* \right] \right)$
	\end{enumerate}
\end{enumerate}	
The lower bound is given by
$$
\textsf{ELBO}(q) = \textsf{E}_q\log p(\yv\mid\Z,\alv,\sig^2) + \textsf{E}_q\log p(\alv,\sig^2) - \textsf{E}_q\log q(\alv) - \textsf{E}_q\log q(\sig^2)\,.
$$
Using standard results, the evaluation of each term is straightforward and results in:
\begin{align*}
\textsf{ELBO}(q) &= -\tfrac{1}{2}(N\log 2\pi + p\log N - p) + a_\sig\log b_\sig - \log\Gamma(a_\sig) \\
&\hspace{0.4cm} + a^*(1 + \log b^* - 2\psi(a^*)) + \log \Gamma(a^*) + 2(\log b^* - \psi(a^*)) + \tfrac12\log|\V^*|\\
&\hspace{0.4cm}-\frac{a^*}{b^*}\left[ b_\sig + \tfrac12 \left( \|\yv\|^2 -2\yv^\trans\Z\mv^* + \mv^{*\trans}\left( \Z^\trans\Z + \tfrac{1}{N}\I_p \right)\mv^* + \tr\left[\left( \Z^\trans\Z + \tfrac{1}{N}\I_p \right) \V^* \right] \right) \right]\,,
\end{align*}
where $\psi(\cdot)$ is the digamma function.

Posterior summaries can be obtained via Monte Carlo simulation using standard random number generation routines.

{\color{black}
\section{Notation}

Matrices and vectors with entries consisting of subscripted variables are denoted by a boldfaced version of the letter for that variable. For example, $\xv = (x_1,\ldots, x_n)$ denotes an $n\times1$ column vector with entries $x_1,\ldots, x_n$. We use $\zerov$ and $\onev$ to denote the column vector with all entries equal to 0 and 1, respectively, and $\I$ to denote the identity matrix. A subindex in this context refers to the corresponding dimension; for instance, $\I_n$ denotes the $n\times n$ identity matrix.
The transpose of a vector $\xv$ is denoted by $\xv^{\top}$; analogously for matrices.
Moreover, if $\X$ is a square matrix, we use $\tr(\X)$ to denote its trace and $\X^{-1}$ to denote its inverse. The norm of $\xv$, given by $\sqrt{\xv^\top\xv}$, is denoted by $\|\xv\|$.
}

\end{document}